%% file: main.tex
\documentclass[preprint,journal]{vgtc}            % preprint (journal style)

\vgtccategory{Research}

\vgtcpapertype{theory/model}

\title{Progressive Value Reading: The Use of Motion to Gradually Examine Data Involving Large Magnitudes}

\author{%
  \authororcid{Leni Yang}{0000-0003-4527-4905},
  \authororcid{Aymeric Ferron}{0009-0001-8881-2729},
  \authororcid{Yvonne Jansen}{0000-0001-5092-551X},
  \authororcid{Pierre Dragicevic}{0000-0002-1854-5899}
}

\authorfooter{
  \item Leni Yang is with Inria, CNRS. E-mail: leni.yang@inria.fr 
  \item Aymeric Ferron is with Inria, CNRS, Univ. Bordeaux. E-mail: aymeric.ferron@inria.fr
  \item Yvonne Jansen is with Univ.~Bordeaux, CNRS, Inria, LaBRI. E-mail: yvonne.jansen@cnrs.fr
  \item Pierre Dragicevic is with Inria, CNRS. E-mail: pierre.dragicevic@inria.fr
}

\abstract{%
People often struggle to interpret data with extremely large or small values, or ranges spanning multiple orders of magnitude. While traditional approaches, such as log scales and multiscale visualizations, can help, we explore in this article a different approach used in some emerging designs: the use of motion to let viewers gradually experience magnitude---for example, interactive graphics that require long scrolling or street paintings stretching hundreds of meters. This approach typically demands substantial time and sustained interaction, translating differences in magnitude into a visceral sense of duration and effort.
\rv{Although largely underexplored, this design strategy offers new opportunities. We introduce the term \textit{\termName} to refer to the use of motion to progressively examine an information object that encodes a value, where the amount of motion reflects the value. We compiled a corpus of \casenum real-life and hypothetical visualization examples that allow, encourage, or require \termName.} From this corpus, we derived a design space of ten design dimensions, providing a shared vocabulary, inspiration for novel techniques, and a foundation for empirical evaluation. An online corpus is also available for exploration.
}


\teaser{
   \centering
   \includegraphics[width=\linewidth, alt={This figure shows portions of a web infographic titled Wealth, Shown to Scale. The infographic takes the form of an extraordinarily long stacked chart, drawn at a fixed scale of \$1,000 per pixel. The top panel in the figure provides an overview of the entire visualization. The overview is The first mark visible to viewers is a single pixel to represent \$1,000. It is followed by larger squares that highlight scale differences between the median yearly U.S. household income, which is \$63,719, \$1 million, and \$1 billion. As viewers continue scrolling to the right, the visualization reveals immense rectangles representing the wealth of Jeff Bezos in 2021, using 278,000 x 500 pixels, and of the richest 400 Americans, using 5,920,000 x 500 pixels.}]{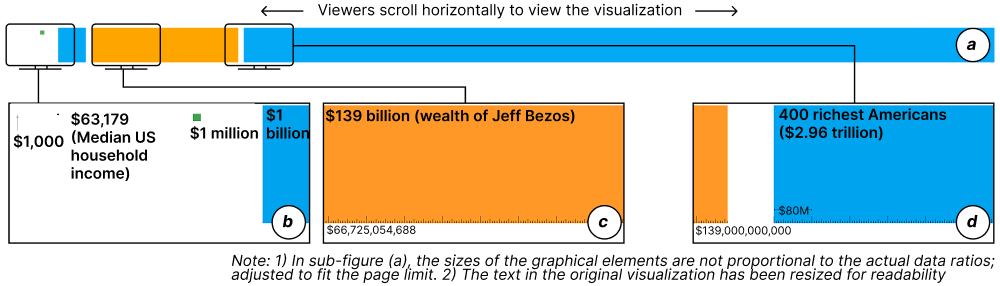}
   \caption{Excerpts from a web infographic titled \emph{``Wealth, Shown to Scale''}~\ecite{C1} (\rv{credit: github user mkorostoff, forked by Luís Lyra}), illustrating wealth inequality in the U.S. The infographic takes the form of an extraordinarily-long horizontal stacked chart, drawn at a fixed scale of \$1{,}000 per pixel, requiring viewers to scroll extensively. The top panel \textit{(a)} provides an overview of the entire visualization. The initial view \textit{(b)} contains a single pixel depicting \$1{,}000, followed by larger squares representing \$63,719 (the median yearly U.S.\ household income), \$1 million, and \$1 billion. As viewers continue scrolling to the right, the visualization reveals immense rectangles representing the wealth of Jeff Bezos in 2021 \textit{(c)} (278,000$\times$500 pixels) and of the richest 400 Americans \textit{(d)} (5,920,000$\times$500 pixels). Scrolling through the entire visualization takes several minutes of uninterrupted scrolling; a counter at the bottom provides feedback how far along one is.}
   \label{fig:wealth}
}

\graphicspath{{figs/}{figures/}{pictures/}{images/}{./}} % where to search for the images

\usepackage{amsmath,amsfonts}
\usepackage{algorithmic}
\usepackage[caption=false,font=normalsize,labelfont=sf,textfont=sf]{subfig}
\usepackage{textcomp}
\usepackage{stfloats}
\usepackage{verbatim}
\usepackage{graphicx}
\def\BibTeX{{\rm B\kern-.05em{\sc i\kern-.025em b}\kern-.08em
    T\kern-.1667em\lower.7ex\hbox{E}\kern-.125emX}}
\usepackage{balance}
\usepackage{enumitem}
\usepackage{xcolor}
\usepackage{comment}
\usepackage{lipsum}
\usepackage{soul}
\usepackage{xspace}
\usepackage{ragged2e}
\usepackage{caption}
\usepackage{wrapfig}
\usepackage{makecell, colortbl, longtable, booktabs, multirow, tabularx, array, changepage, threeparttable} % for table
\usepackage{xurl} 
\usepackage[normalem]{ulem}
\usepackage{xspace}
\newcolumntype{L}{>{\raggedright\arraybackslash}X} % wrap + ragged-right

\AtBeginDocument{

}

\newcommand{\pierrehere}[1]{\textcolor{blue}{---------------- Pierre is currently here in his polishing pass}}
\newcommand{\ecite}[1]{\hyperref[ex:#1]{[#1]}}
\newcommand{\stat}[1]{\textcolor{black}{#1}}
\newcommand{\casenum}[0]{55\xspace}
\newcommand{\realcasenum}[0]{36\xspace}

\newcommand{\rv}[1]{\textcolor{black}{#1}}
\newcolumntype{Y}{>{\RaggedRight\arraybackslash}X}
\newcommand{\termName}{\rv{progressive value reading}\xspace}
\newcommand{\termDesign}{\rv{progressive value reading design}\xspace}
\newcommand{\termDesigns}{\rv{progressive value reading designs}\xspace}
\newcommand{\termNameCap}{\rv{Progressive value reading}\xspace}
\usepackage{mathptmx}                  % use matching math font

\begin{document}

%%%%%%%%%%%%%%%%%%%%%%%%%%%%%%%%%%%%%%%%%%%%%%%%%%%%%%%%%%%%%%%%
%%%%%%%%%%%%%%%%%%%%%% START OF THE PAPER %%%%%%%%%%%%%%%%%%%%%%
%%%%%%%%%%%%%%%%%%%%%%%%%%%%%%%%%%%%%%%%%%%%%%%%%%%%%%%%%%%%%%%%

%% The ``\maketitle'' command must be the first command after the
%% ``\begin{document}'' command. It prepares and prints the title block.
%% the only exception to this rule is the \firstsection command

\maketitle
\input{sections/1-introduction}
\input{sections/2-related_work}
\input{sections/3-method}

\input{sections/4-terminology}

\input{sections/5-design_space}
\input{sections/6-discussion}
\input{sections/7-conclusion}

%% if specified like this the section will be omitted in review mode
\acknowledgments{%
All the simple drawing style figures in our paper were created using ChatGPT. They were later revised by the authors manually.
\rv{We thank all workshop participants, including senior researchers, PhD students, and research engineers of the Bivwac team at Inria.}
}

\bibliographystyle{abbrv-doi-hyperref-narrow}

\bibliography{ref}
\clearpage 
\onecolumn
\appendix
\input{sections/8-appendix}
\twocolumn
\end{document}

%% file: sections/1-introduction.tex
\section{Introduction}

Data involving large magnitudes plays a critical role in understanding pressing societal issues, including government spending~\cite{boyce2022large} and carbon emissions~\cite{batziakoudi2024designing}.  
We use the term \textit{large-magnitude data} broadly to refer to data that features large numbers, often combined with large ratios or differences between values, or ranges spanning multiple orders of magnitude.  
Past research showed that people often have difficulty understanding and interpreting large-magnitude data, which can result in distorted judgments.
For example, studies have shown that people’s valuation of a problem does not rise proportionally with its scope, a phenomenon known as \textit{scope insensitivity}~\cite{carson1997contingent,kahneman1999economic}. People are willing to pay almost the same amount to help 2,000, 20,000, and 200,000 migrating birds from drowning in uncovered oil ponds~\cite{desvousges2010measuring}. Similarly, studies have found that people's empathy does not grow and may even diminish as the number of victims grows, a phenomenon known as \textit{psychic numbing}~\cite{slovic2007if}.

Chart designers have long developed techniques to help people understand large-magnitude data. In particular, log scales have long been used in science and engineering~\cite{brinton1919graphic}. Lately, there has been a renewed interest in strategies for conveying large-magnitude data among visualization designers and researchers. For example, Batziakoudi et al.~\cite{batziakoudi2024lost} recently surveyed techniques for visualizing multiple orders of magnitude by separately encoding mantissae and exponents, while Solen et al.~\cite{solen2024designv1,solen2024designv2} recently contributed a broad survey of multiscale visualization techniques. In addition, unconventional visual encodings, such as data analogies and concrete scales, have been proposed to help general audiences relate complex measures to familiar concepts~\cite{chevalier2013using,chen2024beyond}. 

\begin{figure}[bt]
    \centering
    \includegraphics[width=0.5\textwidth,alt={The figure shows three examples of \termDesigns. The first example is a chart painted on a street. It compares Yale University's annual contribution to its home city of New Haven with its \$32 billion endowment. The contribution is represented by a very small red rectangle. The \$32 billion endowment is represented by a blue 200-meter-long stripe. The second example is a YouTube video called 8 Billion People in Perspective. The video shows the entire world population as a huge crowd assembled in a vast area. Every individual on the Earth is represented by a 3D human model. In the video, the camera moves through the crowd and gradually lifts up. Architectural landmarks are used to give a sense of scale.}]{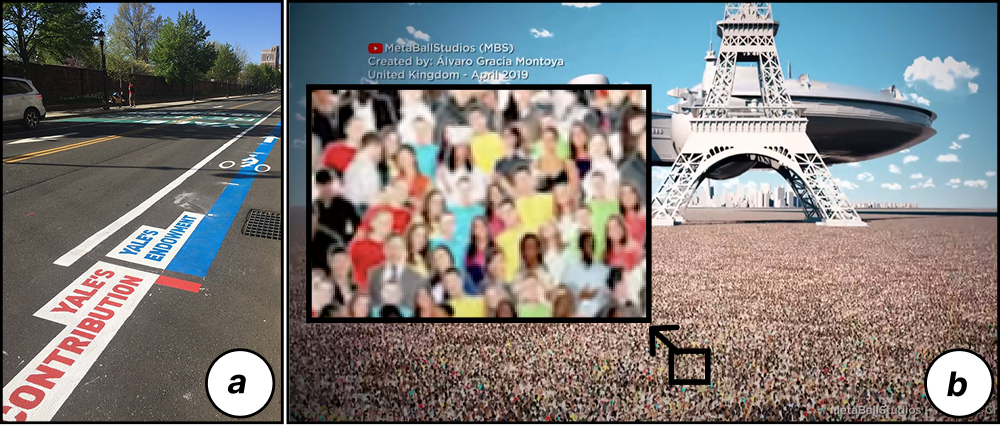}
    \caption{Two examples of \termDesigns: (a) Chart painted on a street showing Yale University's annual contribution to its home city of New Haven (red rectangle) vs. its \$32 billion endowment (blue stripe, 200-meter long)~\ecite{C4} (\rv{Photo credit: Davarian L. Baldwin}). (b) The video \textit{8 Billion People in Perspective} shows the entire world population as a huge crowd assembled in a vast area\rv{, while} architectural landmarks are used to give a sense of scale \rv{\ecite{C2} (image credit: MetaBallStudios).}} 
    \label{fig:intro-cases}
\end{figure}

Past research has explored a variety of techniques for conveying large-magnitude data. However, many possibilities have yet to be fully explored. In particular, we observed emergent designs that encourage viewers to gradually examine individual data values.
For example, \textit{Wealth Shown to Scale}~\ecite{C1} (\cref{fig:wealth}) is an interactive visualization that conveys the wealth of the median US household income, of billionaire Jeff Bezos, and of the richest 400 Americans using a linear scale of \$1,000 per pixel. To view the entire visualization, viewers must scroll horizontally during extended periods of time. 
There exist physical implementations of similar concepts. For example, the \textit{Respect New Haven} activist group painted a 200-meter-long bar chart on a street, requiring viewers to walk or bike along it to fully examine a single value~\ecite{C3} (\cref{fig:intro-cases}-a).
Similar effects can be achieved without requiring any action from the viewer. For example, the animated video \textit{8 Billion People in Perspective}~\ecite{C2} (\cref{fig:intro-cases}-b) depicts the entire world population densely packed across a vast outdoor area, with an overhead camera panning backward during two minutes to reveal the scale of the crowd.

All such examples use motion to let viewers understand large magnitudes progressively.
When sustained motion is required, the experience of time or effort can become a byproduct of the process, augmenting the perception of the magnitude being represented.
The underlying idea is simple but opens up a rich design space. 
Existing designs often incorporate strategies to enhance comprehension, reduce boredom, and maintain engagement. For example, the \textit{8 Billion People in Perspective}~\ecite{C22} video mentioned previously presents models of city buildings and famous landmarks like the Eiffel Tower, scaled to compare with the human crowds (\cref{fig:intro-cases}-b).
The example in \cref{fig:wealth} includes engaging text narratives and tick marks to help viewers estimate the length of the visualization.
% \rvdelete{As another example, the web infographic \textit{500,000 Lives Lost}~\ecite{C22} (\cref{fig:intro-cases}-c) is a large scrollable bee-swarm plot \cite{eklund2016beeswarm} showing COVID-19 deaths over time, and includes text narratives and a mini-overview to help viewers track their position and jump to any point in the visualization.}
Previous literature has touched on related designs, as detailed in \cref{sec:related_work}. However, there is a lack of definitions and a design space to capture the complexity and variety of design options.
 
\rv{To fill this gap, we introduce the concept of \textit{progressive value reading}, which refers to the gradual examination of a visualization element that conveys a single value.}
% We use \textit{\termDesign} to refer to visualization designs that \textit{allow, encourage, or require} \termName. 
%By \textit{information objects}, we mean objects in \textit{data visualizations} that have to be adjusted if the data changes \cite[Chap.3, p.139]{von2002language}, such as 
\rv{Examples include the bars in \cref{fig:wealth}, the painted stripes in \cref{fig:intro-cases}-a, and the 3D crowd in \cref{fig:intro-cases}-b.
\cref{sec:term} goes through terminology and definitions in more detail.}

\rv{We conducted a survey and a workshop}, resulting in a collection of \casenum real and hypothetical examples of \termDesigns. 
Based on these examples, we developed a design space which encompasses 10 design dimensions. 
The design space aims to provide a shared language for communication between designers and researchers, guide design by informing key design elements, inspire future designs by integrating approaches that are often considered separately and revealing unexplored areas, and support the development of evaluation studies by clarifying relevant comparison dimensions.
Our accompanying website\footnote{\label{fn:website}\href{https://progressive-value-reading.github.io/}{https://progressive-value-reading.github.io/}.} allows users to explore our corpus of examples.

%% file: sections/2-related_work.tex
\section{Related Work}
\label{sec:related_work}

We review related work including visualizations for large-magnitude data, motion in visualization, unconventional data encodings, and then focus on work closely related to \termName.

\subsection{Visualizing Large-Magnitude Data}
\label{sec:rela-vis-for-large-data}

When visualizing data spanning multiple orders of magnitude, smaller values typically become illegible. 
Many strategies manipulate visual scales to accommodate both large and small values.
A long-standing and widely used approach is to employ \textit{logarithmic scales}~\cite{brinton1919graphic}.
Another type of approach combines multiple linear scales.
For example, in \textit{dual-scale charts}, one scale spans the entire value range to show the data context, and another scale zooms in to show small values~\cite{isenberg2011study}. 
Borgo et al.~\cite{borgo2014order} introduced \textit{order-of-magnitude markers}, compound visual marks that combine two elements: one that encodes the mantissa\footnote{In scientific notation, a value such as 32,000,000 = 3.2$\times$$10^6$ has mantissa 3.2 and exponent 6.} of the data value, and the other one that encodes the exponent.
Recently, Batziakoudi et al.~\cite{batziakoudi2024lost} presented a comprehensive design space of such visualizations, which separately visualize the mantissa and the exponent of data values. Meanwhile, Solen et al.~\cite{solen2024designv1, solen2024designv2} introduced a design space of \textit{multiscale visualizations}, designed to convey data whose range is too large to be effectively represented on a single scale. A subset of this design space covers closely related concepts, which we discuss in detail later in this section.

Although visualizations that distort, combine, or break down numerical scales can effectively accommodate multiple orders of magnitude, they are typically designed for analytical purposes and technical audiences. When communicating with the general public, however, these techniques can be inaccessible or overwhelming. Consequently, alternative approaches have been developed to engage broader audiences, emphasizing intuitive understanding of large-magnitude data rather than numerical precision or analytical reasoning~\cite{boyce2022large}. In particular, there has been growing interest in the use of \textit{concrete scales}~\cite{chevalier2013using} and \textit{data analogies}~\cite{chen2024beyond} in infographics, where familiar objects (e.g., a huge stack of dollar bills to represent national debt) help people grasp otherwise abstract values. 
As we will see, many \termDesigns also employ concrete scales or data analogies. The design space of techniques discussed in this article differs from the above literature in its focus on \textit{motion} as a means to \textit{progressively} convey large-magnitude data.

\subsection{Motion and Data Visualization}
\label{sec:rw_motion}

There has been a lot of prior work discussing the role of motion in data visualization. A primary purpose of motion is to show data changes over time~\cite{veras2019saliency,robertson2008effectiveness}. 
Motion has also been explored as an alternative visual channel to encode data attributes~\cite{lu2020enhancing,bartram1997perceptual,bartram2002filtering,bartram2003moticons,ware2004motion}, as a means of creating animated transitions between different datasets and visual representations~\cite{chevalier2016animations,heer2007animated}, and for narrative intents~\cite{lan2021kineticharts,shi2021communicating,amini2018hooked}.
Generally speaking, most interactive visualizations involve motion, as many user actions (e.g., brush, pan) result in continuous changes in the visual representation~\cite{dimara2019interaction}.

More directly relevant to our work, Yao et al. \cite{yao2022visualization} surveyed \textit{visualization in motion}, i.e., situations where there is relative motion between an entire visualization and the viewer. They distinguish between three cases: (1) \textit{moving visualization, stationary viewer}, (2) \textit{stationary visualization, moving viewer}, and (3) \textit{moving visualization, moving viewer}. Visualizations that \rv{leverage} \termName \rv{are visualizations} in motion, and can involve any of these three cases. However, the reverse implication does not hold---Yao et al. discuss several examples that are not \termDesigns, such as sports videos augmented with visualizations that move with the players, or visualizations on smartwatches. Yao et al. mentioned one example of \termDesign (the street chart in \autoref{fig:intro-cases}-a) but did not elaborate further on the concept.

Another relevant research stream explores the use of physical locomotion in data visualization, especially in large-display settings.
For example, research has found that locomotion contributes to the benefits of large displays on performance measures, including spatial memory~\cite{ball2007move, jansen2019effects, jakobsen2015moving}. 
A variety of locomotion-based large display visualization designs have been proposed. 
For example, Jakobsen et al.~\cite{jakobsen2013information} proposed proxemics-driven interactions that take the distance, orientation, movement, and location of the users as input to achieve information visualization tasks on large displays.
Isenberg et al.~\cite{isenberg2013hybrid} introduced hybrid-image visualizations, static data representations whose appearance changes depending on the users' distance to the large display. 
Recently, Jain et al.~\cite{jain2025strollytelling} had participants walk through data stories in virtual reality either through physical locomotion (a technique they called \textit{strollytelling}) or through virtual locomotion. They found that participants prefer physical locomotion over virtual locomotion. 
Overall, previous work suggests that physical locomotion may promote engagement and understanding when used in conjunction with visualization. Our design space includes locomotion as a key dimension.

\subsection{Unconventional Data Encodings}

While conventional graphical encodings have long been---and continue to be---the predominant and often most effective way of visualizing data~\cite{card1999readings}, researchers have explored and discussed a range of other methods.
For example, it has been suggested that non-visual sensory modalities such as haptic~\cite{paneels2009review},
auditory~\cite{kaper1999data}, olfactory~\cite{patnaik2018information}, and even gustatory~\cite{hogan2017towards} modalities might complement visual representations. In data physicalizations, data is mapped to the geometry or material properties of physical artifacts, which can be perceived actively and through multiple senses~\cite{jansen2015opportunities,dragicevic2021data}. 
Recent work explores the mapping of data to complex human experiences.
Lee et al.~\cite{lee2020data} introduced \textit{data visceralization} that leverages data-driven virtual reality experiences, particularly for communicating physical measures (e.g., speed, distance, and size). Similarly, Casamayou et al.~\cite{casamayou2022ride} presented \textit{Ride Your Data}, where people ride a roller coaster in virtual reality to experience increases and decreases in data values.
We focus the rest of this section on work where data is mapped to time or effort, as with \termDesigns.

\subsubsection{Mapping Data to Time}

A straightforward way to map data to time is through an animated visualization, where the temporal dimension of the dataset is directly aligned with the passage of time in the animation \cite{lu2020enhancing, robertson2008effectiveness}. For example, a scatterplot of country indicators can be animated to show the evolution of indicators over time~\cite{rosling2011health}. Time has also been used as a channel for encoding non-temporal data, especially within non-visual representations.
In data sonifications, where data is represented through sound, data can be mapped to sound duration~\cite{smith1991data, elmquist2025sensibly}. 
For haptic displays, studies have found that the duration of a vibrotactile stimulus can be effectively used to convey distance information to drivers via a tactile waist belt~\cite{ferguson2018evaluating}, and that people associate increased vibrotactile duration with higher amplitudes of quantities such as size, distance, accuracy, error, and danger~\cite{asif2010exploring}. In all such examples, however, durations tend to be short, and the experience of passing time does not serve to reinforce perceptions of large magnitudes.

\subsubsection{Mapping Data to Effort}

Mapping data to physical effort has been discussed in a few papers, usually as a strategy to raise awareness and promote behavior change.
One study asked participants to ride a stationary bike to generate the amount of energy consumed by a Google search query, and found that this experience fostered greater understanding and engagement than only waiting for a progress bar to fill up~\cite{hurtienne2020move}. In another study, participants were placed in a virtual bathroom and were asked to repeatedly use a water bottle to fill a tank corresponding to a toilet flush or a one-minute shower, which significantly increased their behavioral intentions to conserve water one month later~\cite{hsu2018using}. Another paper speculated that having individuals carry the weight equivalent of their personal carbon emissions during daily activities could increase awareness and understanding~\cite{chauvergne2024weight}. 
Although previous work has suggested that mental effort may be desirable in visualization~\cite{hullman2011benefitting}, we do not know of work specifically exploring the mapping of data values to mental effort. Overall, our work complements these previous explorations by examining a general method for mapping data to effort. However, \termName represents a broader concept, as it may not involve any interaction, effortful or otherwise.

\subsection{Previous Work on Progressive Value Reading}
\label{sec:AROM}

A narrative strategy closely related to \termName is \textit{scrollytelling}, a form of online journalism where a visual story unfolds as the reader scrolls~\cite{seyser2018scrollytelling,tjarnhage2023impact}, and which can be used to convey data~\cite{peng2024telling}. Lan et al.~\cite{lan2022negative} described deliberately increasing visualization length---a strategy they call \textit{stretched layout}---as a means to generate tension in scrollytelling.
Although data scrollytelling stories can involve \termName (and we will discuss several examples), they do not need to. Conversely, \termName can be achieved through methods other than scrolling. 
As mentioned previously, Jain et al. \cite{jain2025strollytelling} introduced \emph{s\textbf{t}rollytelling}, where users navigate data stories presented in virtual reality by strolling along a pre-defined path. However, their designs mainly involved walking through a series of flat displays, with no example involving \termName.

A recent workshop paper by Ferron et al.~\cite{Ferron.etal2025} introduces the concept of AROM---Augmented Ramble along large Orders of Magnitude. AROMs are large-scale AR visualizations that invite people to explore large magnitudes by walking. The paper describes two implemented proofs of concepts: in \textit{ClimbingChart}, a large-scale bar chart is displayed vertically across several floors of a building, while in \textit{WalkingChart}, a large-scale bar chart is displayed horizontally over several meters of a parking lot. In both cases, viewers need to move along a bar to explore and understand the values. Both cases qualify as \termDesigns as per our definition of \autoref{sec:term}, and are included in our corpus. However, the paper only illustrates the concept through two prototypes, without discussing the entire design space.

Closest to our work is a subset of the design space by Solen et al.~\cite{solen2024designv1, solen2024designv2} mentioned earlier. They introduce \textit{visceral time} as a dimension, defined as \textit{``whether a visualization relies on the user's experience of time passing while navigating''}~\cite{solen2024designv1, solen2024designv2}. They further introduce \textit{lengthy pan} as a design strategy consisting of \textit{``a single total scale that the user pans along and which relies on visceral time''}, with five example implementations~\cite{solen2024designv1}.\footnote{Only included in the preprint~\cite{solen2024designv1}; the final published version~\cite{solen2024designv2} omits lengthy pan but retains visceral time.} While lengthy pan closely resembles our notion of \termName and their examples overlap ours, it constitutes only a small, briefly discussed portion of their design space. In contrast, we extend this concept beyond simple panning, exploring it far more comprehensively and in greater detail. We further contrast our design space with Solen et al.\ in \autoref{sec:comparison}.

%% file: sections/3-method.tex
\section{\rv{Method}}
\label{sec:method}
We developed a design space of \termName through an iterative process of (1) compiling a corpus of examples, (2) identifying dimensions and refining terminology, and (3) coding examples to probe and refine the design space dimensions. 
Our corpus includes \casenum examples of which \realcasenum are \textbf{real-life examples}, 13 are \textbf{workshop examples} from a brainstorming workshop, and 6 are \textbf{illustrative examples} created by the authors after the design space stabilized to illustrate some strategies in the paper. For simplicity, we use \textit{hypothetical examples} to refer to both workshop and illustrative examples throughout the paper.

\subsection{Examples Collection and Curation}
\label{sec:case_collection}
A primary challenge was the lack of an established term for a reliable keyword-based search for examples as is common in related work on data storytelling design spaces~\cite{yang2021design,amini2015understanding,shi2021communicating}.  
We therefore used an exploratory case collection strategy inspired by prior work~\cite{morais2020showing,lan2022negative,solen2024designv2}. In addition, we found screen-based articles and videos were dominant in real-life examples, which limit our ability to explore broader design possibilities across different media. Thus, we conducted a brainstorming workshop for more examples. All examples and descriptions are listed in \autoref{sec:appendix} and on the project website~\textsuperscript{\ref{fn:website}}.

\textbf{Real-life examples.}
We collected \realcasenum real-life examples from: 
(1) prior academic corpora and surveys of related techniques~\cite{yang2023understanding,lan2022negative,solen2024designv2,chen2024beyond,chevalier2013using,jakobsen2015moving,lee2020data}; and 
(2) portfolios of creators identified from examples, including both established news agencies (e.g., New York Times and Reuters), and independent designers. 
% We began with a small set of seed examples and collected similar artifacts we encountered. 

\textbf{Workshop examples.}
To expand beyond predominantly screen-based real-life examples, we ran a structured brainstorming workshop with 15 researchers external to the project but from the same lab. They were senior researchers, PhD students, and research engineers with backgrounds in immersive analytics, data visualization, and HCI. We started the workshop by introducing our initial working definition and some examples. We ended up with 13 workshop examples. 

To structure ideation and ensure coverage of heterogeneous media, we provided three scaffolding dimensions: \textit{Display}, \textit{Viewer locomotion mode}. and  \textit{Body movement}. We chose these because they capture major constraints on how \termName can occur across media (screen, immersive, physical) and how users traverse information. Participants were split into 5 groups. Each group ideated for 40 mins on two or three assigned combinations from the \stat{12} combinations of \textit{Display} $\times$ \textit{Viewer locomotion mode}. The dimension \textit{Body movement} was left unconstrained to avoid prematurely narrowing interaction possibilities.
Each group presented concepts and provided feedback, which we used to revise the term definition and identify additional candidate dimensions.

\textbf{Illustrative examples.}
The design space was ultimately stabilized through cross-case coding across the corpus including the real-life and workshop examples, with real-life examples serving as the primary grounding. 
During design space development, 6 examples were brainstormed by the authors to verify if some categories in a dimension or certain combinations of dimensions are possible. 
They are used for verification and illustration not for major design space development.

\subsection{Design Space Development}
\label{sec:design_space_method}
We developed the design space using a combination of top-down and bottom-up methods.
From the top-down perspective, we began from general representation and interaction dimensions commonly used in visualization and interaction taxonomies (e.g., \textit{Display}, \textit{Body movement})~\cite{jakobsen2015moving,solen2024designv1}. These served as a starting vocabulary, not a fixed schema. From the bottom-up perspective, we iteratively coded examples to extract recurring strategies and refine or add dimensions (e.g., \textit{Progress Review}, \textit{Anchor}, \textit{Narrative}). 
% When recurring patterns in our data matched established concepts, we used prior work (e.g., [30, 33]) to ground our strategy definitions and terminology.

After the first version of the design space with 9 candidate dimensions, we iteratively refined it over three rounds.\footnote{All iterations are documented in supplementary materials on OSF \label{fn:osf_link}\url{https://osf.io/pxn82/overview}}) First, all authors coded \stat{8} diverse examples to surface overlooked strategies and propose modifications. This expanded the design space from \stat{9} to \stat{11} candidate dimensions. Next, we agreed a tentative definition for each of the dimensions, which then the first and second author used to code the collected corpus (\stat{44} cases at that time). Based on the coded corpus, we sharpened our definitions and agreed to remove \stat{2} redundant (always coded identically) or ambiguous dimensions (low agreement between co-authors) and added a new one, resulting in \stat{10} dimensions. In parallel to the design space development, we continued to grow our corpus, for example, by snowballing from authors of examples already included in our corpus and by brainstorming among co-authors to generate illustrative examples for as of yet unexplored combinations of design dimensions, which resulted in \stat{11} additional examples. With each addition we verified that the ten dimensions were sufficient to describe them without introducing new dimensions or requiring restructuring, thus suggesting the design space is stabilized. Finally, the first author re-coded all \stat{55} examples.

%% file: sections/4-terminology.tex
\section{\rv{Basic Terminology}}
\label{sec:term}

We define \textit{progressive value reading} as:

\vspace{-1mm}
\begin{quote}
the use of motion to progressively examine an \rv{information object} that encodes a value, where the amount of motion reflects the value.
\end{quote}
\vspace{-1mm}
\termNameCap \rv{therefore} refers to a \textit{user activity}, which may be active (e.g., panning, walking) or passive (e.g., watching a video).
By extension, \rv{we use the term \textit{\termDesign} to refer to a visualization design that allows, encourages, or requires \termName.} \rv{Note that the concept is not related to progressive visualization, which refers to visualizations that incrementally update with incoming data \cite{ulmer2023survey}}. 

Our definition of progressive value reading is deliberately broad and does not necessarily require the examination of large-magnitude data. However, in this paper we focus on cases involving large magnitudes, as this is the most common application in our corpus. Additional uses are discussed in \cref{sec:whenuse}.

Next, we \rv{unpack the definition by discussing} the notion of \textit{information object} and \textit{the use of motion}.

\subsection{Information Objects}
\label{sec:info_object}

\rv{Following Engelhardt's framework \cite{von2002language}, we use \emph{information object} to refer to an element in a visualization that encodes data.\footnote{Engelhardt defines information objects as \emph{``the graphic objects that would have to be adjusted if the information (data) that one intends to represent would change''} \cite[Chap.~3, p.~129]{von2002language}.}
Information objects can be graphical or physical in the case of data physicalizations.
Furthermore, information objects can encode complex information (an entire visualization can be considered an information object \cite{von2002language}), but in this article we focus on information objects that encode a single quantitative value.}

\rv{A typical example of an information object is a \emph{mark}, which is a primitive graphical element that represents a single data item \cite[Chap.~5]{munzner2014visualization}, such as a bar in a bar chart. However, an information object can also be a collection of marks that, together, encode a single quantitative value as is commonly the case for unit charts\cite{drucker2015unifying}.
For instance, \autoref{fig:intro-cases}-b shows a large crowd where each person (mark) represents an individual in the world. Despite the many marks, the infographic's focus of communication is not on the individual, but on the population as a whole. Therefore, the relevant information object is the entire crowd, which conveys a single value (the world population) through the area it takes up (or equivalently, its cardinality). Since this information object is not visible all at once, it requires progressive examination to be read.}
%Similarly, \autoref{fig:case-unit} shows a large-scale unit visualization \cite{drucker2015unifying} where each icon represents an individual in the world, colored according to their continent of residence~\ecite{C16}.
% The area of each collection of marks encodes the population of one continent and is progressively examined.

\begin{figure}[h]
    \centering
    \includegraphics[width=\linewidth, alt={This figure presents a hiking trail. It features a scale model of the solar system where every meter corresponds to one million kilometers of actual distance. The trail starts with the sun, at a size of 1.39 meters, and continues along the planets until Pluto. Each segment of the trail between two planets is progressively examined.}]{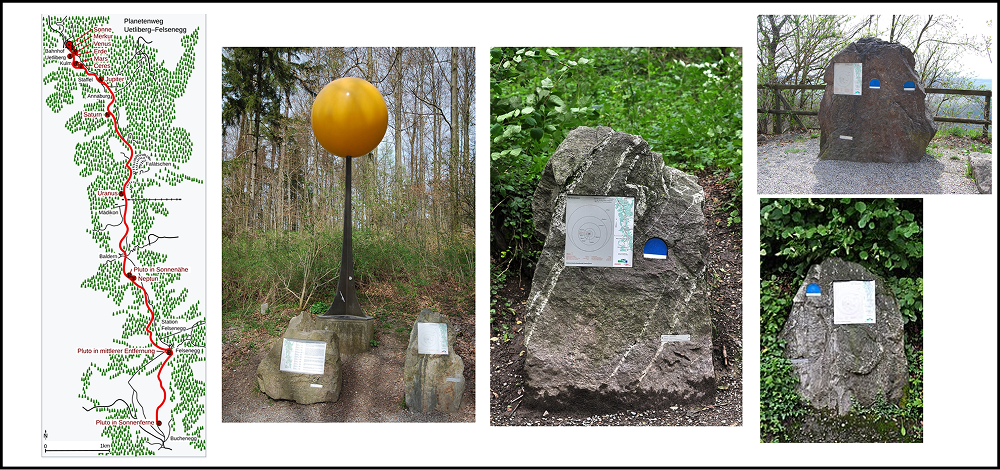}
    \caption{\textit{Planetenweg Uetliberg–Felsenegg}~\ecite{C7} is a hiking trail featuring a scale model of the solar system where every meter corresponds to one million kilometers of actual distance. The trail starts with the sun, at a size of 1.39~m, and continues along the planets until Pluto. Each segment of the trail between two planets is progressively examined. \rv{Photos released under CC0.}}
    \label{fig:case-hike}
    \vspace{-2mm}
\end{figure}

\rv{Finally, an information object can also refer to a physical space without an explicit visual representation.}
For example, \textit{Planetenweg Uetliberg–Felsenegg}~\ecite{C7} (\autoref{fig:case-hike}) is a Swiss hiking route featuring a scale model of the solar system between the Sun and Pluto, where each meter represents one million kilometers.
We consider that the length of each segment of the trail between two planets is \rv{an information object that can be} progressively perceived by hiking along it.

\subsection{The Use of Motion}
\label{sec:motion-define}
By \textit{the use of motion}, we mean any relative movement between the viewer and the \rv{information object}.
Examples include cases where the \rv{information object} moves while the viewer is stationary (such as during scrolling; see~\autoref{fig:wealth}) and cases where it is the other way around (such as during physical locomotion; see~\autoref{fig:case-hike}).
Relative motion can be along the \rv{information object} (e.g., during panning) or toward/away from it (e.g., during zooming).
It can also occur when an \rv{information object} is constructed progressively, such as when many pieces accumulate to form a bar. 

\begin{figure}[h]
    \centering
    \includegraphics[width=\linewidth, alt={This figure presents the screenshot of a YouTube video called Number of Deaths in World War II per Country. In this video, the camera moves through a series of 3D bars. The figure shows the 3D bar chart filmed by the camera from a top view.}]{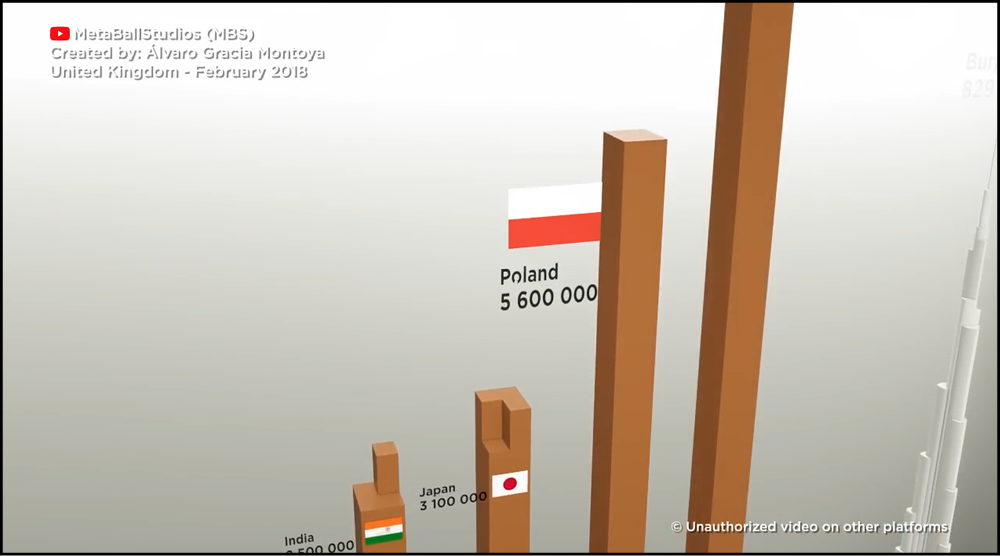}
    \caption{\textit{Number of Deaths in World War II per Country}~\ecite{C24} (Image credit: MetaBallStudios) is a YouTube video in which the camera moves through a series of 3D bars. \termNameCap occurs when the camera moves up to reveal very tall bars.}
    \label{fig:case-ww2}
\end{figure}

For motion to qualify as \termName, the total amount of motion must be roughly proportional to the data value, or at least increase with larger values.
Counterexamples are a bubble chart where the circles drop into position quickly purely for visual effect (lack of proportional relationship between amount of motion and data) or a bar chart that shows many data points on its x-axis, which requires panning along that same axis to see all bars (motion along \textit{different} information objects while each is visible in its entirety).
Similarly, in the video shown in \autoref{fig:case-ww2}, the camera initially pans horizontally over a series of 3D bars. \termNameCap only occurs at the end, when vertical panning is needed to examine the taller bars representing very large death tolls. We include such examples in our corpus as long as they feature at least one instance of \termName.

It is required that motions should be perceptible to the viewer. A counterexample is panning a uniformly colored bar across a uniform background, which produces no visible effect. To make the motion perceptible, additional visual cues must be introduced, such as tick marks (\autoref{fig:wealth}-c,d) or visual patterns (\autoref{fig:intro-cases}-b). In physical environments (e.g., \autoref{fig:intro-cases}-a, \autoref{fig:case-hike}), such cues are typically inherent, and the issue rarely arises.

%% file: sections/5-design_space.tex
\section{Design Space}
\label{sec:design}

\begin{figure*}
    \centering
    \includegraphics[width=\linewidth, alt={This figure is the design space overview. It lists the design space dimensions from top to down. The dimensions are grouped into three groups. How is data represented? How is motion achieved? And, what are the design strategies for engagement and understanding?. Under each dimension, the design space overview gives icons for each design category. All dimensions and categories are explained in the paper.}]{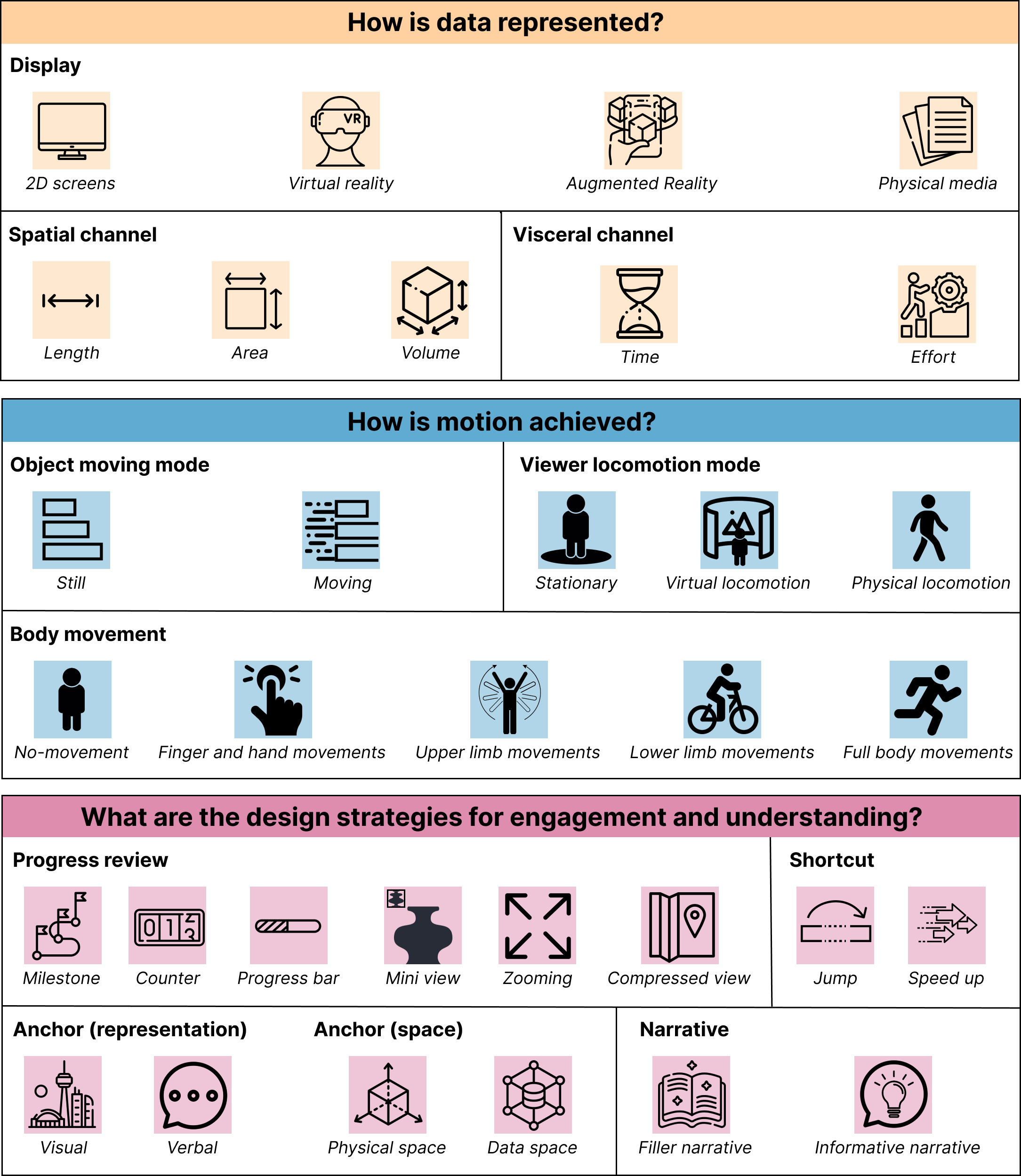}
    \caption{The design space overview.}
    \label{fig:overview}
\end{figure*}

An overview of our design space is shown in~\cref{fig:overview}. The dimensions of the design space can be classified into three groups: \textit{(i)} how is data represented, \textit{(ii)} how is motion achieved, and \textit{(iii)} what are the design strategies for engagement and understanding. We review them in turn. For simplicity, we use the term ``hypothetical example'' to refer to either a workshop or an illustrative example, as discussed in \cref{sec:method}.

\subsection{How is Data Represented?}
This category captures how data is encoded into conventional or unconventional representations, and includes the dimensions \textit{display}, \textit{spatial channel}, and \textit{visceral channel}.

The inclusion of the \textit{display} dimension follows precedents in design spaces of narrative \cite{segel2010narrative} and affective \cite{lan2023affective} visualization, where visualizations span multiple media such as films, comic strips, or interactive apps. It highlights that \termDesigns are not limited to 2D screens and encourages consideration of alternative media such as AR and VR, which offer distinct affordances and technical possibilities. The \textit{spatial channel} dimension builds on Munzner’s taxonomy of visual channels~\cite[Chap.~5]{munzner2014visualization}.
We use the term ``spatial channel'' rather than ``visual channel'' because the encodings \rv{that can support} \termName are restricted to spatial dimensions (length, area, or volume, see \autoref{sec:spatial_channel}), and marks need not be visually rendered (see \autoref{sec:info_object}). Finally, the \textit{visceral channel} dimension builds on the notion of data visceralization~\cite{lee2020data} and reflects a distinctive feature of \termDesigns---their use of time and effort to convey data. We now discuss these three dimensions in more detail.

\subsubsection{Display}

The \textbf{display} dimension captures the type of devices or medium that is used to present the \rv{information objects}.

\textbf{2D screens} include mobile phones, desktop monitors, and wall-sized screens that display \rv{information objects} on a two-dimensional surface. 
This display type allows visualization designs to be shared very broadly, and is the most common in our corpus, with \stat{31 out of \realcasenum} real-life examples. 

\begin{figure}
    \centering
    \includegraphics[width=\linewidth, alt={This figure shows two hypothetical examples. The first example applies virtual reality. The viewers experience a roller coaster with the track made of concatenated stacks of 100-dollar bills. The length of the track represents the wealth of the richest person in the world. The second example applies AR to overlay stacks of 100-dollar bills onto a street. It compares the wealth of normal people and the richest people.}]{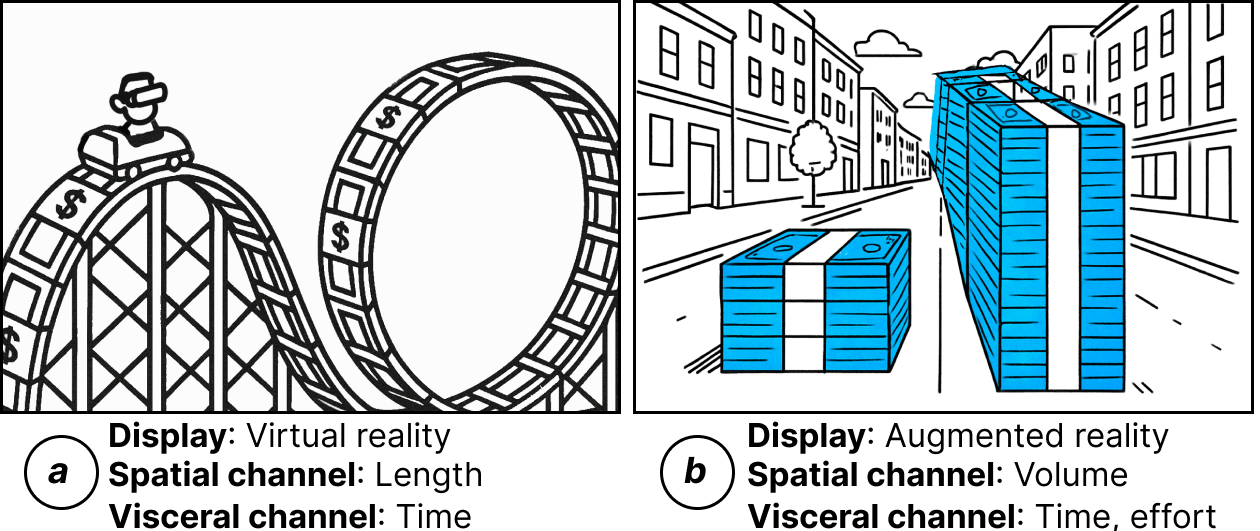}
    \caption{The hypothetical example~\ecite{C40} in sub-figure (a) applies VR in which viewers experience a roller coaster with the track made of concatenated stacks of 100-dollar bills representing the wealth of the richest person in the world. The hypothetical example~\ecite{C42} in sub-figure (b) applies AR to overlay stacks of 100-dollar bills onto a street to compare the wealth of normal and the richest people.}
    \label{fig:display}
\end{figure}

\textbf{Virtual reality (VR)} displays, including VR headsets and CAVE systems, can be used to visualize data, for example, to create a visceral experience~\cite{lee2020data}. \autoref{fig:display}-a illustrates a hypothetical example where a viewer experiences taking a roller coaster in VR with the trails consisting of concatenated stacks of 100-dollar bills to represent the wealth of the richest person in the world. This example is inspired by the work of Casamayou et al.~\cite{casamayou2022ride}.

\textbf{Augmented reality (AR)} displays include AR headsets, glasses, projectors, smartphones, and holographic displays.
These technologies enable the placement of \rv{information objects} within real-world environments, generally at a lower cost than physical representations.
For example, \autoref{fig:display}-b illustrates a hypothetical example~\ecite{C40} where a virtual depiction of stacks of money is overlaid onto a street to compare median wealth with the richest people. Compared to a VR version, this AR installation allows viewers to use their familiar surroundings as comparison anchors~\cite{Ferron.etal2025}.

\textbf{Physical media} encompass a wide range of non-digital formats, such as paper print-outs, paintings, and data physicalizations~\cite{jansen2015opportunities,dragicevic2021data}. An example we mentioned earlier is the 200-meter-long strip painted on a street (\autoref{fig:intro-cases}-a). Physical media can be effective at engaging local audiences.

While 2D screens are currently predominant in our corpus, AR, VR, and physical media offer considerable potential. Since \termDesigns are often for very large values, these alternative displays can leverage the expanded spatial affordances offered by large-scale virtual and physical environments. VR displays are particularly well-suited for integrating \termName into data-driven experiences that cannot be easily realized in the physical world, such as riding a data-driven roller coaster (\autoref{fig:display}-a) or examining planet-scale data visualizations from a spaceship (\autoref{fig:moving}-c). As for physical media, they enable engagement within shared spaces and support creative use of everyday objects, such as using rolls of toilet paper as \rv{information objects} (\autoref{fig:moving}-a) or streetcars as visualization canvases (\autoref{fig:moving}-b).

\subsubsection{Spatial Channel}
\label{sec:spatial_channel}

The \textbf{spatial channel} dimension captures the spatial attribute of the \rv{information object} that is used to encode the data value examined during \termName. 
In our corpus, we identified the use of \textbf{length}, \textbf{area}, and \textbf{volume} as encoding attributes. 
% Length and area are the most frequently used spatial channels, with volume being relatively rare (only 6 out of \realcasenum real-life examples).

% When an information object is a surface, such as a rectangle, it can be unclear whether the spatial channel is length or area. We consider the spatial channel to be length if the information object's size varies along a single dimension as a function of data, such as the strip painted on a street (\autoref{fig:intro-cases}-a). % Conversely, if the information object's size varies along both dimensions, as in the infographic showing the wealth of the richest Americans (\autoref{fig:wealth}), the spatial channel is area. Similarly, for a 3D information object, the spatial channel can be length, area, or volume, depending on how many spatial dimensions are allowed to vary. For example, in \autoref{fig:case-ww2}, the stacks of coffins resemble a bar chart from afar, but individual stacks vary in width and depth and are often irregularly shaped, so we consider the spatial channel to be volume. Note that even for a single-information object visualization like \autoref{fig:intro-cases}-b, the information object can be thought to vary with data if its size \textit{would} have differed had another value been visualized.

\rv{The motion required to progressively examine an \rv{information object} depends on the spatial channel used. When the channel is length, one-dimensional motion (e.g., scrolling) is sufficient. In contrast, fully examining an area or volume requires following a 2D or 3D path. 
For example,  \textit{7 billion People on 1 Page}~\ecite{C16} is a web infographic showing the world population by using human icons to represent every person in the world, where viewers \rv{need to} scroll vertically \textit{and} horizontally to view the entire visualization.
However, \rv{information objects} need not be fully examined for viewers to infer their size: the area of a large circle or square, for instance, can be estimated by traversing its diameter or by zooming in or out. 
Finally, an \rv{information object} may encode values along two or three dimensions but support motion only along certain ones. 
For example, for the tall stacks of money in AR (\autoref{fig:display}-b), viewers can walk around to inspect the lengths of the bill stacks but cannot do the same for their heights.}

Finally, while there exist other encoding channels than length, area, and volume~\cite[Chap. 5]{munzner2014visualization}, many channels are not suitable for \termName. For example, color, shape, and angle are not graphical attributes that can be progressively examined by having the \rv{information object} move relative to the viewer. 

\subsubsection{Visceral Channel}
\label{sec:visceral_channel}

The \textbf{visceral channel} dimension captures whether \textbf{time} or \textbf{effort} is used as an implicit encoding channel to convey the data value, in addition to the spatial channel.

To qualify as a channel, time or effort needs to \textit{(i)} be consciously experienced by the viewer; \textit{(ii)} increase with the data value. By \textit{consciously experienced}, we mean that the time or effort involved in the \termName must be sufficiently intense or prolonged to be subjectively felt by the viewer. 
A counterexample would be a bar chart that only requires a brief scroll to be viewed fully---so brief that it blends with other user interactions and goes mostly unnoticed. \textit{By increases with data value}, we mean that the duration or effort experienced by the viewer grows if the value is larger.\footnote{Again, even with a single value as in \autoref{fig:intro-cases}-b, one can imagine how much time would have been required if the value were larger.} A counterexample would be a bar chart where each bar takes a minute to render because of a progressive calculation \cite{fekete2016progressive}; while the waiting is long enough to be subjectively felt by the user, it is unrelated to the data values conveyed.
These criteria imply that while all \termDesigns involve the use of motion to gradually experience the spatial properties of \rv{information objects} (\autoref{sec:term}), not all cases involve time or effort as meaningful channels.

\textbf{Time} can be used as a visceral channel by mapping large values to large-scale \rv{information objects} and requiring viewers to progressively examine them to perceive their full scale. Examples include the long bars in the scrollable infographic from \autoref{fig:wealth}, and the long bars in the walkable AR visualization \textit{Walkingchart} \ecite{C31}. If the process lasts long enough for the passage of time to be consciously felt, it satisfies our first condition for a visceral channel.
Since the total motion increases with data value (a condition of \termName, see \autoref{sec:info_object}), if the motion proceeds at an approximately constant speed, then time itself increases with data value, satisfying our second condition for a visceral channel.

\textbf{Effort} can be used as a visceral channel also by mapping large values to large-scale \rv{information objects}, and this time requiring effortful movements from viewers. Examples include a long, continuous scrolling on a website, or physical activities such as walking (as in the street-painted chart in \autoref{fig:display}-1) or hiking (as in the solar system hike in \autoref{fig:case-hike}). An even more striking example is walking upstairs, as in \textit{ClimbingChart} \ecite{C30}.
Again, if the physical effort is salient enough to be consciously experienced and total effort increases with data value, it satisfies the conditions for a visceral channel.

Whether time or effort qualifies as a visceral channel may depend on the user. The same amount of waiting or effort may be highly salient to some but may go unnoticed by others. Moreover, when users can adjust the motion speed, the proportionality between data value and time or effort may be disrupted. For simplicity, we consider a visualization to support a visceral channel if it likely meets the necessary conditions for most users in typical use cases.

\subsection{How is Motion Achieved?}
\label{sec:motion}

This category captures how the motion is realized, and includes the dimensions \textit{object moving mode}, \textit{viewer locomotion mode}, and \textit{body movement}.
These three dimensions capture the core component of \termName---the use of motion. The \textit{object moving mode} and \textit{viewer locomotion mode} specify the nature of the motion, extending the typology described by Yao et al.~\cite{yao2022visualization} (see \autoref{sec:rw_motion}). The \textit{body movement} dimension further characterizes the actions required to sustain that motion.

\subsubsection{Object Moving Mode} 

\begin{figure}
    \centering
    \includegraphics[width=\linewidth,alt={This figure shows four examples. The first example uses a toilet paper roll to compare the sizes of different digital media. The second example paints the bars of carbon emissions of different modes of transportation on the side of a tram. The third example is a spaceship journey in virtual reality. The Earth is surrounded by stacks of 100-dollar bills in VR. It represents the wealth of the richest person. The last example uses a phone as a peephole view. The viewer holds the phone to inspect the visualization hidden on a wall.}]{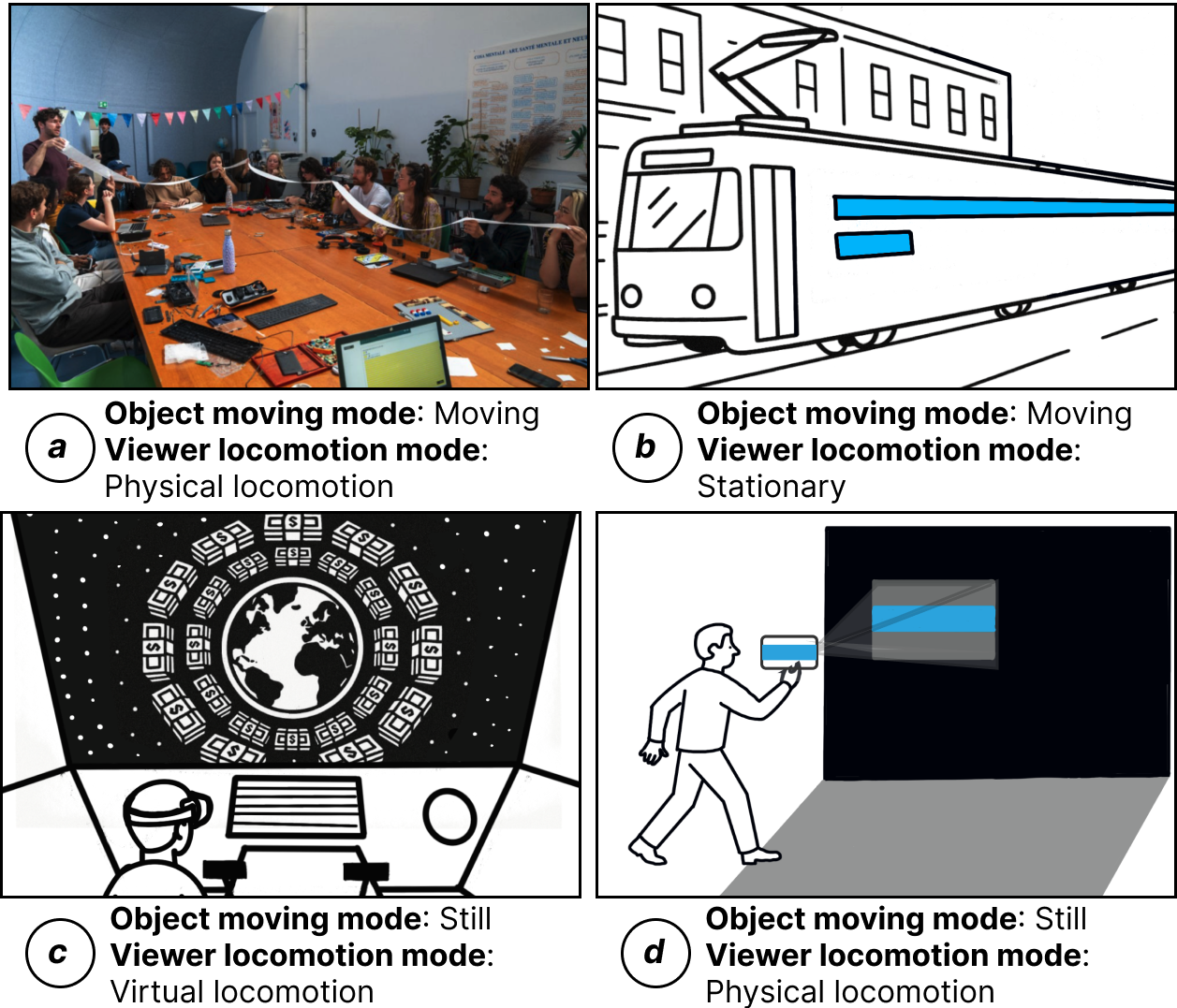}
    \caption{Different combinations of object moving modes and viewer locomotion modes: (a) The use of a toilet paper roll to compare sizes of different digital media~\ecite{C17} (\rv{Photo credit: Antoine Aphesbero}). The example is special, as the viewers could be invited to unroll the paper or sit and observe. We used it to show how viewers and \rv{information objects} can move together; (b) A hypothetical example~\ecite{C51} in which the carbon emissions of different modes of transportation are painted on the side of a tram; (c) A hypothetical example~\ecite{C44} in which viewers take a virtual spaceship journey around the Earth that is surrounded by stacks of 100-dollar bills representing the wealth of the richest person; and (d) A hypothetical example~\ecite{C39} that uses a phone as a peephole view through which the viewer can inspect the visualization on a wall.}
    \label{fig:moving}
\end{figure}

\textbf{Object moving mode} refers to whether there are changes in the position of the \rv{information object} or a build-up process during the \termName process. 

An \rv{information object} is \textbf{still} when there is no build-up process and its position is fixed---or appears fixed---throughout the \termName process. For example, the trail segments between planets in the hiking path are fixed (\autoref{fig:case-hike}). For digital media, we consider that an object is fixed if the user \textit{perceives} it as such, which includes the VR rollercoaster and the AR dollar stacks (\autoref{fig:display}), as well as the coffin stacks in the 3D data video (\autoref{fig:case-ww2}). 

An \rv{information object} is \textbf{moving} when there is a build-up process or its position changes---or appears to change---during the \termName process. A typical example is when viewers scroll through a large-scale visualization on the webpage and the \rv{information objects} move into the viewport (\autoref{fig:wealth}). \rv{Information objects} can also move in much larger physical spaces. For instance, in the hypothetical example~\ecite{C51} of \autoref{fig:moving}-b, carbon emissions from different modes of transportation are visualized along the body of a tram. As the tram moves through the city, passersby observe the data when it travels past them. This example is inspired by statistical graphics presented in wagons in the Municipal Parade by the employees of the City of New York in 1913~\cite[Page.342]{brinton1919graphic}.
An example of a build-up process is in~\autoref{fig:moving}-a. Specifically, the pieces of a paper roll represent the sizes of various digital media (e.g.\ an email, a website, a high-resolution photo). To see the size of a Netflix movie, viewers need to unroll the paper roll to reveal the \rv{information object} gradually.  

\subsubsection{Viewer Locomotion Mode}
\label{sec:viewer_locomotion_mode}

\textbf{Viewer locomotion mode} refers to whether the viewer's body position changes or appears to change during \termName. The viewer locomotion mode is \textbf{stationary} when viewers neither change their physical location nor perceive any locomotion.

The viewer locomotion mode is \textbf{virtual locomotion} when viewers perceive that their body is moving through space despite remaining physically stationary---a perceptual phenomenon also known as \textit{vection} \cite{palmisano2015future}.
VR is probably the most prevalent and accessible technique to achieve virtual locomotion. \autoref{fig:moving}-c illustrates a hypothetical example~\ecite{C44} where viewers embark on a virtual spaceship journey around the Earth, which is surrounded by stacks of 100-dollar bills representing the wealth of the richest individual.

The viewer locomotion mode is \textbf{physical locomotion} when the viewer’s body actually changes position during \termName, such as while walking or during vehicular locomotion. Physical locomotion allows inspection of large-scale \rv{information objects} that vanish into the distance, whether in the real world, as the 200-meter-long strip painted on the street (\autoref{fig:intro-cases}-a), or in virtual environments. \autoref{fig:moving}-d illustrates a different scenario, where viewers move along a wall and use a phone as a peephole view through which the viewer can inspect the hidden \rv{information objects}. 

While both virtual and physical locomotion can provide the subjective experience of movement, they involve different trade-offs. Virtual locomotion requires more technical expertise to implement, but it can be more practical to deploy and does not require viewers to travel to a specific location. However, physical locomotion---particularly walking---has been shown to improve people’s ability to estimate distances~\cite{wiesing2023serial}, which can help viewers perceive the scale of large \rv{information objects}. Achieving a comparable level of spatial understanding in VR generally requires both physical motion (e.g., walking on a treadmill or in a large tracking space~\cite{ruddle2011walking}) and realistic visual cues~\cite{wiesing2023serial}.

\subsubsection{Body Movements}

\begin{figure*}
    \centering
    \includegraphics[width=\textwidth,alt={This figure shows three examples. They include several combinations of viewer locomotion modes and body movements. In the first example, the viewers take an elevator with a transparent window. Through the window, they can see a bar chart painted on the wall. The second example is an animated timeline visualization of the universe. The timeline goes forward as the viewer turns the crank. In the third example, the viewers ride a stationary bike to circle around down-scaled planets in virtual reality. It is for comparing their sizes.}]{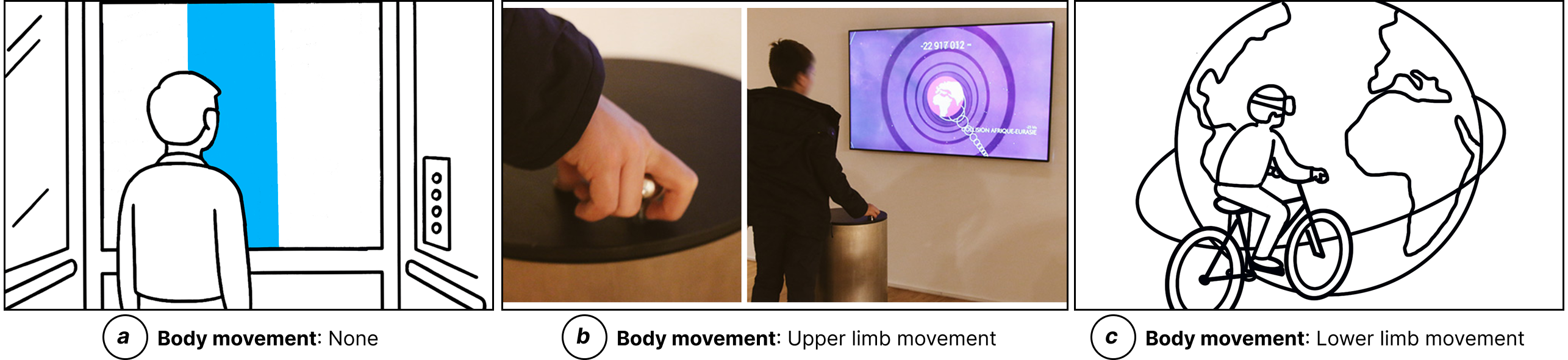}
    \caption{Several combinations of viewer locomotion modes and body movements: (a) A hypothetical example~\ecite{C54} where viewers take an elevator with a transparent window through which they can see a bar chart painted on the wall; (b) The animated timeline visualization of the universe goes forward as the viewer turns the crank~\ecite{C5} (\rv{Photo credit: Fleur de Papier}); (c) A hypothetical example~\ecite{C45} where viewers ride a stationary bike to circle around down-scaled planets in virtual reality to compare their sizes.}
    \label{fig:movement}
\end{figure*}

The \textbf{body movement} dimension captures the movements performed by the viewer to \textit{sustain} \termName. It does not include the body movements only needed to trigger, pause, or stop the process, such as clicking the play button of a video. The categories of this dimension are based on which body parts drive the movements. For each example, we coded which body parts provided essential force or control. For example, riding a bicycle is classified as a full-body movement because it requires coordinated use of both upper and lower limbs for balance, steering, and propulsion. In contrast, riding a stationary bicycle is coded as a lower-limb movement, since the legs supply the primary effort and the upper limbs are not essential.

In the \textbf{no-movement} case, the viewer does not need to take any action to sustain the motion. A typical example is watching a video.
Note that it is possible to rely on physical locomotion while no body movement is required to sustain \termName. For example, in the hypothetical example~\ecite{C54} illustrated in~\autoref{fig:movement}-a, the viewer takes an elevator with a transparent window, through which they can see a long bar chart painted on the wall that is gradually revealed as the elevator moves up. This example is inspired by a 1985 art piece by Keith Haring \cite{debur2005Haring}.

\textbf{Finger and hand movements} are the most common movement in our corpus (\stat{21 out of \realcasenum} real-life example). Most examples are online articles with scrolling interaction.  

\textbf{Upper limb movements} involve the use of arms. An example is illustrated in~\autoref{fig:moving}-b, where the viewer is required to turn a crank. 

\textbf{Lower limb movements} involve the use of legs, with the exclusion of other body parts. For example, in~\autoref{fig:movement}-c, the viewer rides a stationary bike to circle around different planet models in VR to compare their sizes. 

\textbf{Full-body movements} need the coordination of the whole body, such as during walking (\autoref{fig:intro-cases}-a and~\autoref{fig:case-hike}). 

Requiring body movements to sustain \termName is a straightforward way to use effort as a visceral channel.
In addition, bodily experiences and interactions with the environment are known to shape and support cognition~\cite{clark1999embodied}. Studies show that task-relevant movements aid learning in mathematics~\cite{duijzer2019embodied}; for example, children who trace temperature graphs with their finger perform better on transfer tests~\cite{agostinho2015giving}. By extension, \termDesigns that engage the viewer's body may support a deeper understanding of data.

\begin{figure}
    \centering
    \includegraphics[width=\linewidth, alt={This figure shows a virtual reality experience. The viewer virtually descends into the sea along a freediving line.}]{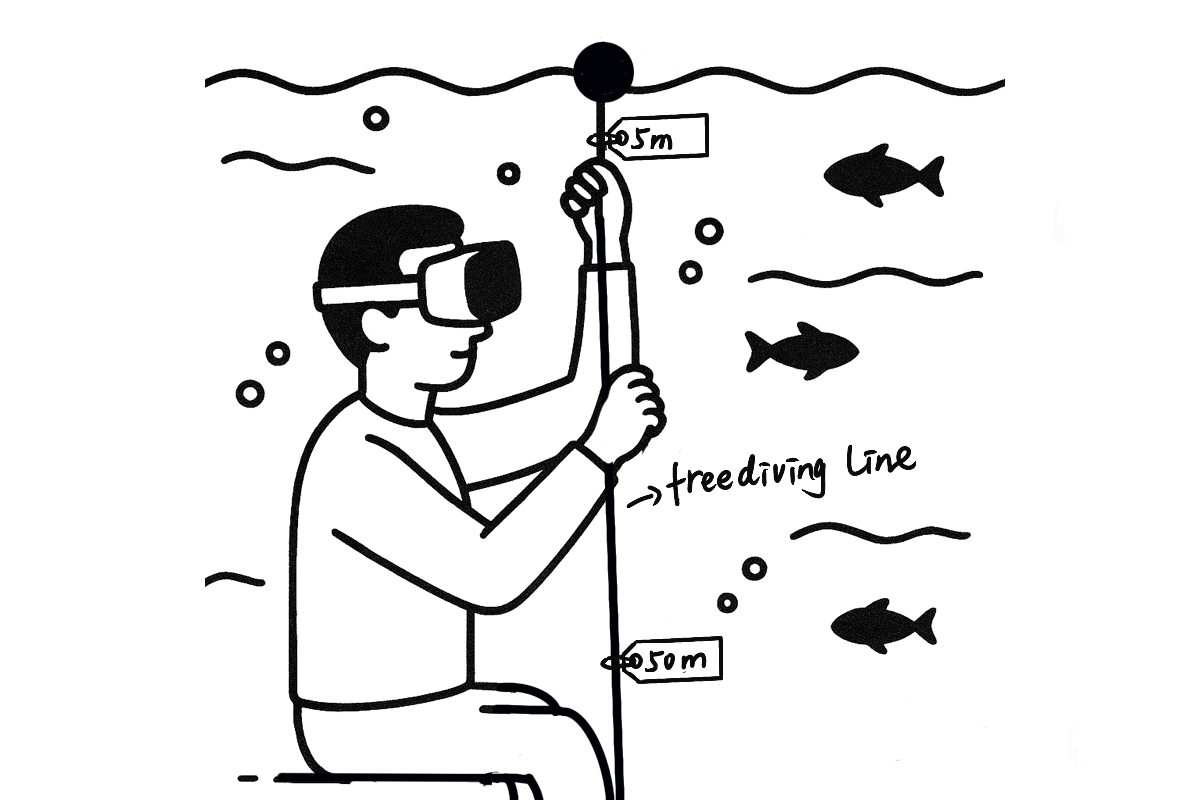}
    \caption{A hypothetical \textit{virtual locomotion} adaptation \ecite{C55} of the \textit{The Deep Sea} infographic \ecite{C10}, in which the viewer virtually descends into the sea along a freediving line.}
    \label{fig:case-diver}
\end{figure}

Many interesting combinations of \textit{viewer locomotion mode} (\autoref{sec:viewer_locomotion_mode}) and \textit{body movement} remain to be explored, including accessibility--oriented or recreational setups (e.g., wheelchairs, handcars, or treadmills). Designs using one combination can often be adapted to another: a stationary-viewer scrollytelling piece can be turned into a virtual-locomotion experience by embedding \rv{information objects} in a 3D world. Such adaptations may require rethinking body movements, as in example~\ecite{C55} (\autoref{fig:case-diver}), which reimagines \textit{The Deep Sea} (\autoref{fig:progress}-a) using freediving gestures. Adapting designs to physical media can pose practical challenges, but installations such as the physical re-creation of \textit{Wealth Shown to Scale}~\ecite{C29} show that walkable variants can offer engaging experiences in public or museum settings.

\subsection{What Are the Design Strategies for Engagement and Understanding?}
\label{sec:strategies_engagement}

\termNameCap can be lengthy and possibly lead to boredom from repetition or disorientation when viewers lose track of their position or progress. 
This category covers design strategies to reduce these issues, including the dimensions \textit{progress review}, \textit{shortcut}, \textit{anchor}, and \textit{narrative}.
Several of these strategies have appeared in prior research, but here we emphasize those particularly relevant when \termName becomes lengthy or effortful. For example, mechanisms for reviewing progress can reduce disorientation; shortcuts can give viewers more control over the experience; anchors can support the perception of scale in large-magnitude data; and narrative elements can increase engagement. We now discuss each dimension in detail.
% These strategies address challenges common to \termName---such as uncertainty about one’s position, repetitive motion, or difficulty perceiving scale---and can help maintain engagement and comprehension. For example, mechanisms for reviewing progress can reduce disorientation; shortcuts can give viewers more control over the experience; anchors can support the perception of scale in large-magnitude data; and narrative elements can provide structure and context. We now discuss each dimension in detail.

\subsubsection{Progress Review}
\label{sec:progress_review}
\textbf{Progress review} consists of showing viewers how far they are from the beginning or the end of the full \rv{information object}. We list possible techniques, based on Cockburn et al.'s \cite{cockburn2009review} survey of focus+context techniques as well as our own corpus.

\begin{figure}
    \centering
    \includegraphics[width=\linewidth, alt={This figure shows three examples. The first example The Deep Sea. It is a visualization of ocean depth with animals shown at different levels. The second example is an illustration. The past and future portions of a large \rv{information object} in three-dimensional space are folded into a zig-zag path. The third example is What If I Told You: You Eat 3496 Litres of Water. It is a visualization of average daily water consumption. It presents data through a wall of bottles. Those bottles can exactly contain the amount of water.}]{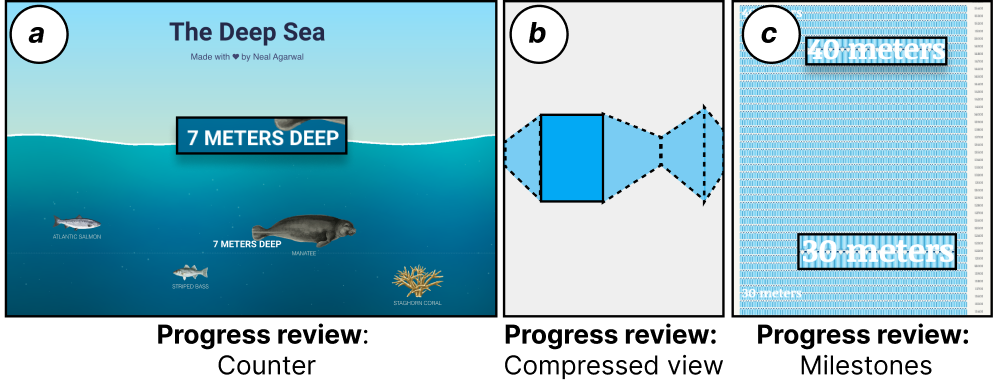}
    \caption{Examples of progress review designs: (a) \textit{The Deep Sea}~\ecite{C10} (\rv{Image credit: Neal Agarwal}), a visualization showing how deep the sea is and animals living at different depth levels; (b) An illustration where the already-visited and upcoming portions of a large \rv{information object} in a 3D space are folded in a zig-zag path. (c) \textit{What If I Told You: You Eat 3496 Litres of Water}~\ecite{C21} (\rv{Image credit: InfoDesignLab}) that discusses the average daily water consumption of a person, considering water not only for domestic usage but also for producing products.}
    \label{fig:progress}
\end{figure} 

A \textbf{progress bar} indicates how far viewers have progressed. In web articles and infographics employing scrollytelling, the default web browser's scroll bar can partly serve this function---although it reflects progress only across the entire web page.
% , not relative to a specific \termName operation.

A \textbf{mini view} is a miniature presentation of the whole \rv{information object} with a position indicator, as in example~\ecite{C22}.
In user interfaces, this type of design is also known as overview+detail~\cite{cockburn2009review}.

A \textbf{counter} is a dynamic numeric display showing the current data value. In~\autoref{fig:progress}-a, viewers scroll through a visualization of the sea’s depth and its animals, while text dynamically updates to show their current depth.

\textbf{Milestones} are discrete indications of how much has been completed or how far viewers are from the end, analogous to kilometer markers on highways. Unlike progress bars and counters, milestones do not dynamically update. As an example, \autoref{fig:progress}-c shows a wall of plastic bottles representing water consumption, with milestones every 10 meters.

\textbf{Zooming} enables viewers to zoom in and out to smoothly transition between detailed views and an overview of an \rv{information object}~\cite{cockburn2009review}. Zooming can be either user-controlled or animated; As an example, in the data video of showing the world population as a huge crowd assembled (\autoref{fig:case-ww2}), the camera pulls back at the end to reveal an overview of all people. 

A \textbf{compressed view} is an alternative visual representation of upcoming or already-visited \rv{information objects} (or \rv{information object} portions) that takes up less space. 
The paper roll in example in~\autoref{fig:moving}-a can be considered to have compressed views both of upcoming \rv{information objects} for the unrolled part and also once unrolled because it is folded along the room to fit a smaller space. 
Compressed views can use various metaphors, such as optical lenses~\cite{cockburn2009review}, folding and perspective~\cite{elmqvist2009melange,mackinlay1991perspective}, or sedimentation~\cite{huron:hal-00846260}. \autoref{fig:progress}-b shows an illustration where a large \rv{information object} stands in a 3D space, and the already-visited and upcoming portions of the \rv{information object} are folded in a zig-zag path. 

Overall, progress reviews can serve three purposes: indicating when the viewer will finish, showing how much they have completed, and providing a sense of scale. A wide range of techniques is possible and has been published in the HCI literature~\cite{cockburn2009review}, but these have so far not yet been exploited in the context of \termName.

\subsubsection{Shortcut}
A shortcut allows the viewer to skip parts of the \rv{information object} or speed up the process of examining the \rv{information object}.

By \textbf{jumping}, viewers can move directly to specific positions of an \rv{information object}. For instance, in the example with a mini view at the top left of the visualization~\ecite{C22}, the viewer can click any position in the mini view to jump to the corresponding part of the article. Direct jumps can cause viewers to become disoriented or lose their sense of scale. This can be mitigated by using animated transitions~\cite{chevalier2016animations} or by incorporating a progress review (\autoref{sec:progress_review}).

By \textbf{speeding up}, a viewer can accelerate the process without disrupting its continuity. Speeding up can be supported either through interactions that give the viewer full control of the motion (e.g., in scrollytelling) or by enabling acceleration of an automated motion (e.g., by adjusting video playback speed).

\subsubsection{Anchor}
\label{sec:anchor}

\begin{figure}
    \centering
    \includegraphics[width=\linewidth, alt={This figure shows four examples combining anchor representations and spaces. The first example is a verbal anchor in physical space from If the Moon Were One Pixel. It says: ``You would need 886 of these screens lined up side by side to show this whole map at once.'' The second example is a verbal anchor in data space from the same example. It says: ``If you were on a road trip, driving at 75km per hour, it would have taken you over 500 years to get here from the earth.''The third example is a visual anchor in physical space from an example titled What If I Told You: You Eat 3496 Litres of Water. It is a cutout of a woman who stands in front of a wall of plastic bottles. The fourth example is a visual anchor in data space from another example titled 500,000 Lives Lost. It puts an image of a train next to a cloud of dots representing the death number. A text explains that the people who died can fill the train.}]{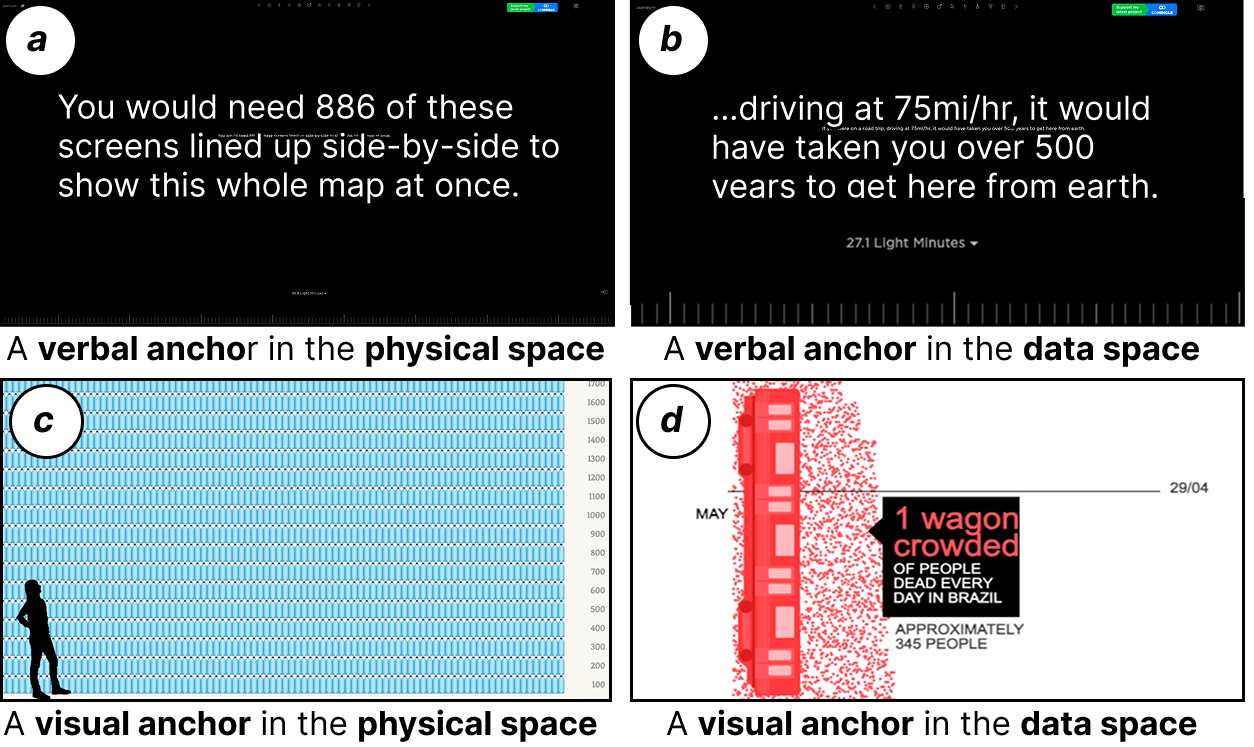}
    \caption{This figure shows the combinations of anchor representations and anchor spaces: (a) A verbal anchor in the physical space from the example \textit{If the Moon Were One Pixel}~\ecite{C19} (\rv{Image credit: Josh Worth}); (b) A verbal anchor in the data space in the example~\ecite{C19}. (c) A visual anchor in the physical space from the example~\ecite{C21}. (d) A visual anchor in the data space from the example~\ecite{C20} (\rv{Image credit: Nexo Journal}).}
    \label{fig:anchor}
\end{figure}

An \textbf{anchor} is a reference point with which the viewer can compare the scale of an \rv{information object} or the magnitude of the value it represents. Anchors can be classified along two sub-dimensions:

The first sub-dimension is \textbf{anchor representation}, which divides anchors into (1) \textbf{visual anchors}, i.e., graphical representations of concepts (\autoref{fig:anchor}--bottom); and (2) \textbf{verbal anchors}, i.e., pieces of text and recorded speech (\autoref{fig:anchor}--top). Visual anchors are closely related to visualizations using concrete scales \cite{chevalier2013using}. 

The second sub-dimension is \textbf{anchor space}, which divides anchors into (1) \textbf{physical-space anchors}, which provide a baseline of comparison for the \rv{information object}'s spatial encoding properties (e.g., length, area, or volume; see \autoref{sec:spatial_channel}); and (2) \textbf{data-space anchors}, which provide a baseline of comparison for the data value conveyed by the \rv{information object}.

All combinations of the two sub-dimensions are possible and illustrated in \autoref{fig:anchor} with real-world examples from our corpus. Figures \ref{fig:anchor}-a and \ref{fig:anchor}-b come from an online article that invites readers to scroll horizontally through the solar system, where the moon is scaled to a single pixel on the screen. In \autoref{fig:anchor}-a, a \textit{verbal, physical-space anchor} explains how many screens are needed to display the entire visualization. Meanwhile, \autoref{fig:anchor}-b presents a \textit{verbal, data-space anchor} that compares planetary distances to the time it would take to cover them by car.
\autoref{fig:anchor}-c provides a \textit{visual, physical-space anchor}, showing the size of a human next to a wall of plastic bottles representing water consumption. Finally, \autoref{fig:anchor}-d illustrates a \textit{visual, data-space anchor}, where a train wagon is used to compare its passenger capacity with the daily number of Covid-related deaths in Brazil at that time.

\subsubsection{Narrative}

A \textbf{narrative} is verbal information that is presented during \termName. 

An \textbf{informative narrative} conveys additional information and messages about the data and its context. For example, the web infographics \textit{Wealth Shown to Scale} (\autoref{fig:wealth}) uses text to narrate what wealth could achieve if it were not in the hands of a single person. %Informing narratives are often combined with other design techniques. By definition,

A \textbf{filler narrative} does not introduce new information; its main purpose is to break the monotony of examining \rv{information objects} and to sustain the viewer’s interest. For instance, the example~\ecite{C19} features a long scroll through an empty, dark space (\autoref{fig:anchor}-a,b). Along the way, sentences such as \textit{“pretty empty out here”} and \textit{“as it turns out, things are pretty far apart”} appear, serving to make the experience less monotonous.

Narratives overlap and interact with several of the dimensions previously discussed. In particular, \textit{verbal anchors} (\autoref{sec:anchor}) are a special type of informative narrative, and \textit{milestones} (\autoref{sec:progress_review}) can be accompanied by narratives that help highlight important moments or transitions in a data story.

% Although narratives are intended to enhance engagement, they must be carefully designed, or they can have the opposite effect. In the example~\ecite{C19} (\autoref{fig:anchor}-a,b), the very long \rv{information object} requires fast scrolling, which causes text messages to appear too briefly, forcing viewers to scroll back only to find they are meaningless fillers. During the coding process, we found that the narratives in this example disrupted rather than enhanced our experience. By contrast, the example of wealth data (\autoref{fig:wealth}) demonstrates a more polished design, incorporating informative narratives with carefully timed and positioned text that remains visible long enough to be easily read.

\subsection{Five Strategies to Achieve Motion}
\label{sec:general}
We now go through five common design strategies to achieve motion that are used in examples from our corpus. These five strategies are not necessarily exhaustive nor exclusive. For example, a visualization inviting the viewer to inspect a very long physical \rv{information object} by driving a car would involve both \textit{A. Interactive panning} and \textit{D. Vehicular motion}.

\textbf{A. Interactive panning.} 
\textit{Interactive panning} refers to designs where only part of a large \rv{information object} is visible through a viewport, and sustained viewer actions shift it so new portions gradually enter and exit the viewport. In 19 examples, this movement uses a scroll wheel or trackpad (i.e., finger and hand movement). It can, but rarely, be produced by other bodily actions, such as turning a wheel (e.g., \autoref{fig:movement}-b). 
Interactive panning generally involves a \textit{moving object} and a \textit{stationary viewer}, but in some designs such as \autoref{fig:progress}-a, some viewers may experience the \rv{information object} as fixed and the viewer moving.

\textbf{B. Animated panning.} Like interactive panning, animated panning brings regions of \rv{information objects} into and out of the viewport, but the motion is predefined, as in the data video illustrated in \autoref{fig:intro-cases}-b. 
This strategy typically involves a \textit{moving object} and a \textit{stationary viewer} but can also feature a \textit{still object} with \textit{virtual or physical locomotion}, as in the case~\ecite{C12}, where the viewer’s camera pans over a large 3D chart in VR.

\rv{\textbf{C. Walkable \rv{information objects}.}} Walkable \rv{information objects} are large-scale \rv{information objects} spread across a vast space, where viewers walk to inspect them fully, as in \autoref{fig:intro-cases}a and \autoref{fig:case-hike}. 
In contrast to panning, walkable \rv{information objects} are not necessarily bounded by explicitly defined viewports, nor are they limited to predefined paths. However, a viewer can still inspect a walkable \rv{information object} by moving sideways in a manner similar to panning. Walkable \rv{information objects} allow viewers to compare scale against the surrounding environment, while naturally integrating full-body movement.

\textbf{D. Vehicular motion.} This strategy uses vehicles (e.g., trains, cars, or elevators---either virtual or physical) to move either \rv{information objects} or viewers. It combines \textit{moving object} with \textit{stationary viewer}, or \textit{still object} with \textit{virtual/physical locomotion}.

\rv{\textbf{E. \rv{Information object} build-up.}} As discussed in~\autoref{sec:motion}, \rv{information objects} can be gradually constructed during \termName. When animated on a screen (e.g., in a data video), this approach combines a \textit{moving object} with a \textit{stationary viewer}. \rv{Information object} build-up can also involve a moving observer, as in \autoref{fig:moving}-a, which combines \textit{moving object} with \textit{physical locomotion}.

\subsection{Excluded Dimensions}
\label{sec:dimension-iteration}
As mentioned in \autoref{sec:design_space_method}, we considered but eventually excluded two dimensions from our design space.

\textbf{\rv{Information object} type}. We initially attempted to categorize \rv{information objects} based on the graphical primitive they employ (e.g., point, line, area), for which several visualization grammars provide typologies~\cite{wilkinson2011grammar,satyanarayan2016vega,munzner2014visualization}. However, visualization designs are becoming increasingly diverse, frequently incorporating more complex forms such as compound \rv{information objects}~\cite{senay1990rules,borgo2013glyph}, pictograms~\cite{haroz2015isotype} with varying degrees of visual realism~\cite{morais2020showing}, and physical \rv{information objects}~\cite{jansen2015opportunities}. To our knowledge, no single, up-to-date framework captures the full range of existing \rv{information object} types. As creating such a framework is beyond the scope of this paper, we have chosen to exclude this dimension.

\textbf{Motion characteristics}. We initially considered including a set of dimensions to describe motion, such as speed, duration, and the direction or dimensionality of motion trajectories (briefly discussed in \autoref{sec:spatial_channel}). However, these aspects proved difficult to assess in our examples, in part because they are often determined by the viewer. For instance, viewers have full control over speed and duration in scrollytelling, as well as over their walking trajectories in physical environments. Therefore, we chose not to include this dimension.

\subsection{Differences with Solen and Colleagues}
\label{sec:comparison}

\newcommand{\solenDS}{DS{\footnotesize\textsc{mul}}\xspace}
\newcommand{\ourDS}{DS{\footnotesize\textsc{pvr}}\xspace}

In the sections above, we described how dimensions and categories previously explored in other contexts have been adapted or reinterpreted within our design space. Considering the design space as a whole, the work of Solen et al.~\cite{solen2024designv1,solen2024designv2} (hereafter referred to as \solenDS) is most closely related to ours (hereafter referred to as \ourDS), and we therefore contrast the two in detail.

Although both design spaces address the visualization of large-magnitude data, they differ in focus. \solenDS{} targets multiscale visualizations, where the range between the largest and smallest values is too wide to be displayed at a single scale. It focuses on how to visually display \textit{different scales} and how to help viewers navigate between them. In contrast, \ourDS{} emphasizes techniques that allow viewers to progressively examine \textit{individual values}.

As a result, the two design spaces differ in the types of datasets they cover. While both cover datasets with multiple orders of magnitudes and large differences in magnitude, \ourDS{} also covers datasets involving a single large value (e.g., \autoref{fig:intro-cases}-b). Furthermore, combining and navigating multiple scales is central in \solenDS{}, but it only plays a peripheral role in \ourDS{}, where focus+context and zooming techniques serve mainly to keep viewers oriented.
%  during \termName.
% (e.g., \autoref{fig:intro-cases}-c). 
Conversely, \ourDS{} thoroughly examines \termDesigns, whereas the concept appears only briefly in the first version of \solenDS{} \cite{solen2024designv1} and is nearly absent in the second \cite{solen2024designv2}.

Specifically, the \solenDS{} dimensions that overlap with \ourDS{} are:

\begin{itemize}[topsep=0pt,itemsep=0pt,partopsep=0pt,parsep=0pt,leftmargin=*]
\item \textbf{Navigation mode.} \solenDS{} distinguishes between \textit{digital} and \textit{physical} navigation modes. The digital mode roughly maps to \textit{digital media + still viewer + finger and hand movements + moving object} in \ourDS, while the physical mode covers other examples, including \textit{physical objects} and \textit{physical locomotion}. \ourDS significantly refines this concept by distinguishing multiple categories of \textit{display}, \textit{object movement}, \textit{viewer locomotion}, and \textit{body movement}, enabling higher descriptive precision and coverage of a broader range of examples, including VR experiences and mobile physical objects.
    
\item \textbf{Visceral time.} This binary dimension captures \textit{``whether a visualization relies on the user’s experience of time passing while navigating''}, which aligns well with \textit{time as a visceral channel} in \ourDS. The initial version of \solenDS included five examples combining \textit{length as a spatial channel} and \textit{time as a visceral channel} in \ourDS{}'s terminology---a design pattern named \textit{lengthy pan} in \solenDS{}. \ourDS{} also includes those examples, but additionally covers \textit{effort as a visceral channel}, and \textit{area} and \textit{volume} as alternative spatial channels with which \termName is possible.

\item \textbf{Familiarity.} This binary dimension indicates whether a visualization uses familiar objects to help viewers understand unfamiliar units. It corresponds to the \textit{anchor} dimension in \ourDS, which further distinguishes between \textit{verbal} and \textit{visual} anchors, as well as \textit{physical-space} and \textit{data-space} anchors.
\end{itemize}

In addition, \ourDS{} addresses design strategies to maintain user engagement and enhance understanding during \termName (e.g., \textit{progress reviews}, \textit{shortcuts}, and \textit{narratives}), which \solenDS{} does not address. Yet many multiscale visualization and navigation strategies from \solenDS{} can effectively complement \termDesigns. For example, in the web infographic \textit{Sea Level Rise by the Numbers} \ecite{C23}, each pictogram first represents the volume of a single swimming pool, then aggregates to 1,000 pools, and finally to 1 million. This progressive scaling avoids excessive scrolling and is well captured by \solenDS{}. Therefore, the two design spaces are highly complementary and can be used together.

%% file: sections/6-discussion.tex
\section{Discussion}
\label{sec:discussion}
This section discusses trade-offs with \termDesigns that we have not covered yet, open research questions, and limitations of our conceptual framework.

\subsection{When and When Not to Use Progressive Value Reading}
\label{sec:whenuse}
In their creative use of motion, time, and effort, visualizations that support \termName have the potential to enhance people’s perception of large-magnitude data, stimulate curiosity, and raise awareness on important topics. However, when applied inappropriately, it can easily feel tedious or inefficient, potentially leading to disengagement. Consequently, designers must carefully consider when---and when not---to employ it. While empirical studies will be needed to carve out where it is most appropriate, we offer initial considerations informed by our corpus.

Because \termDesigns artificially increase time and/or effort, it is \textit{not suitable for most analytical tasks}---such as those requiring rapid comparisons or the identification of trends in large datasets---where efficiency and the ability to interpret complex data are essential \cite{batziakoudi2026beyond}. Therefore, it is more suited for:

\textbf{Communicating with non-experts.} Many examples in our corpus are intended to engage non-expert audiences in casual contexts \cite{pousman2007casual}, where efficiency is less critical than in analytical tasks. For example, in museums \ecite{C5} or art exhibitions \ecite{C29}, visitors often arrive with curiosity and a willingness to invest time in immersive, reflective experiences. In such settings, \termDesigns can support prolonged engagement with data stories, as illustrated in \cref{fig:movement}-b. Public awareness campaigns \ecite{C20} and grassroots communication in everyday spaces \ecite{C4} can also leverage it to draw attention to significant issues involving large-magnitude data. For instance, similar to \ecite{C5} (\cref{fig:intro-cases}-a), which featured a long bar chart painted along a street, placing large-scale visualizations in public environments can convey impactful messages without disrupting people’s daily activities. Finally, \termDesigns may prove useful in educational settings \ecite{C17}, particularly when combined with embodied learning \cite{kosmas2019implementing}, in order to help students understand and internalize concepts such as orders of magnitude.

\textbf{Conveying a small number of data points.} \termNameCap designs are the most suitable when the intent is to focus on only a few data points. When applied to more complex data, it can become lengthy and confusing. Many of our examples, for instance, present just one or two values, although additional data is commonly weaved in through narratives, or other design strategies (e.g., \ecite{C1} in \cref{fig:wealth}). Such cases encourage careful reflection on important statistics that might otherwise be overlooked in conventional charts or tables. While traditional representations are easy to process, they can give the false impression that the numbers are fully understood at a glance. In contrast, \termDesigns slow reading but promotes deeper engagement, supporting broader communicative goals~\cite{yang2025making,adar2020communicative, lee2022affective}. This aligns with prior research suggesting that visualization designs sometimes benefit from increasing cognitive load through visual embellishment or by prioritizing interpretive depth over strict perceptual accuracy~\cite{hullman2011benefitting, Correll2014Brain, bertini2020shouldn}.

\textbf{Beyond large-magnitude data.} 
As our many examples illustrate, \termDesigns are particularly suited for conveying large-magnitude values and large-magnitude differences at a visceral level. However, our definition does not require large-magnitude data; it only involves the progressive examination of an \rv{information object} (\cref{sec:term}). For instance, even in a bar chart where one bar is merely twice as tall as the others, a designer might let the smaller bars fill the viewport and require scrolling to reveal the taller bar. As in broken-axis designs~\cite{borgo2014order}, this strategy can improve the legibility of smaller values and prevent a single large value from compressing the entire chart. Beyond improving legibility, it can also be used to emphasize large values or convey the actual size of objects--for example, by requiring users to scroll to see recent increases in pandemic deaths or life expectancy, or by presenting fish species at a 1:1 scale that forces scrolling to view the largest ones. In such cases, the data may not span multiple orders of magnitude
% , yet \termName still functions as a storytelling device to emphasize large values. 
Because this design choice imposes extra interaction and may cause confusion, its benefits must outweigh its costs, underscoring the need for empirical evaluation.

\subsection{Studying Progressive Value Reading}

Based on our corpus, \termDesigns are currently used by infographic designers and communicators, presumably because of its expected benefits, but no empirical investigations into its potential benefits and limits exist. In particular, understanding whether---and to what extent---it can help people understand or relate to large-magnitude data requires empirical studies.

\subsubsection{Testing Potential Benefits}
\label{sec:benefit}
At the most basic level, psychophysics experiments (which we will further discuss in \cref{sec:psychophysics}) can help us understand how \termName affects graphical perception.
Beyond perceptual questions are cognitive questions related to data understanding: for example, does the experience of \termName increase---or possibly lower---viewers' confidence that they understand the data? Can it serves as a cognitive anchor with a lasting influence on how people interpret similar data in the future? There are also higher-level cognitive and behavioral dimensions worth exploring. For example, can \termName be used to amplify a message's emotional impact---such as conveying a sense of urgency when presenting climate crisis data or eliciting empathy when communicating large numbers of victims? 
Can it lead to meaningful shifts in attitude, beliefs, memory, or even long-term behavioral change? Can it evoke experiences strong enough to subsequently induce involuntary memories? \cite{petiot2025effect} There is a rich space of research questions to address in order to better understand its communicative potential.

\subsubsection{Understanding the Limits}
A key limitation of \termDesigns is that the experience can quickly become tedious; if this leads people to abandon the process, its benefits may be entirely lost. 
Prior work~\cite{lan2022negative} reported that intentionally increasing a visualization’s length to require scrolling may elicit negative feelings.
It is therefore important to study when---based on factors such as dataset size or motion duration---such designs cease to be worthwhile because people are likely to give up. This threshold probably varies depending on the context, whether in casual web browsing, educational settings, or controlled studies. Additionally, \termDesigns remain unconventional choices, and many existing infographics, such as \cref{fig:wealth}, likely rely on novelty, humor, and playfulness to engage viewers. Over time, however, these effects might diminish. Could the widespread use of \termDesigns by designers lead to user fatigue and eventual avoidance of such designs, or could this be mitigated by integrating it into richer storytelling?

\subsubsection{Perceptual Investigations}
\label{sec:psychophysics} 
It will be especially important to study the fundamental perceptual mechanisms of \termName. There is a rich literature on graphical perception covering many different visual variables \cite{cleveland_experiment_1986, legge_efficiency_1989, hollands_judgments_1992, talbot_four_2014} and also work on the use of motion in data visualization (as reviewed in \cref{sec:rw_motion}), yet we are not aware of any study that investigated the psychophysics of graphical perception where motion is required to fully perceive the size of an \rv{information object}. Consequently, it is currently unclear how \termName changes the perception of data values compared with visualizations that are fully visible at once.
As it always implies a combination of mapping data to graphical elements, which are explored over time and may or may not require effort, it is difficult to make predictions based on existing research on graphical perception. In addition, research on the graphical perception of large-magnitude data is still sparse \cite{borgo2014order,batziakoudi2026beyond}. Past research on numerosity perception \cite{dehaene_number_1997} and magnitude perception \cite{leibovich_sense_2017} has studied how humans form internal mental representations of magnitude, but most studies focus on comparatively small numbers, up to the range of hundreds at most. There is evidence for a logarithmic mental number line \cite{izard_calibrating_2008} but also for a tendency to underestimate larger magnitudes. More studies are needed to investigate in how far \termName may be able to alleviate the underestimation of large magnitudes and under which combinations of dimensions of our design space. The perceptual studies can also draw on past research adopting an enactive approach to perception, which sees perception as an interplay between action and sensory input \cite{noe2004action}. For example, past research on VR locomotion suggests that the specifics of locomotion implementation can influence how people perceive their walking speed in VR, which may in turn affect their perception of magnitudes when exploring them through locomotion \cite{banton_perception_2005}.

\subsection{Future Directions}

\subsubsection{Calibrating the Experience of Progressive Value Reading}
\label{sec:optimize}
While perceptual experiments can reveal how people generally perceive values through \termName, tailoring designs to viewers' subjective experiences of time and effort remains challenging. Perceptions vary widely across individuals and contexts---a short hike may be effortless for a healthy adult but taxing for others. Time and effort also interact in complex ways. 
For instance, past studies have shown that greater physical load can lead to longer perceived durations of time~\cite{block2016physical}, while conversely, longer time intervals can increase perceived effort~\cite{faulkner2008rating}. Existing empirical data---e.g., from studies comparing perceived exertion across different activity types \cite{borg1987perceived,green2004heart}---could help calibrate \termDesigns for specific populations. 
New experiments could be conducted to refine the data for such applications, e.g., by measuring subjective effort using standard scales \cite{lopes2022perceived} and/or capturing biometric responses, such as heart rate and blood lactate, which are known to correlate with perceived exertion \cite{zinoubi2018relationships}.

Another research direction is to explore \termDesigns that adapt to individual users, building on work in adaptive user interfaces \cite{norcio1989adaptive} and, more recently, user-adaptive visualizations \cite{yanez2025state}. For example, in a web infographic, scrolling sensitivity could be automatically adjusted to match the viewer’s physical ability, available time, or patience. Similarly, anchors, narratives, or shortcuts might be dynamically displayed if fatigue is detected. A user's state could be assessed through specialized equipment, such as eye tracking or neurophysiological sensors, or simply inferred from interaction patterns---for example, a pause in scrolling may indicate disengagement or fatigue.

\subsubsection{Exploring Other Visceral Channels}
In our discussion of visceral channels (\cref{sec:visceral_channel}), we focused on time and physical effort as potential channels. \textit{Mental effort} may also be important to consider. 
Mental effort is often described as the subjective experience of engaging in cognitively demanding tasks, such as those that require suppressing distractions or overriding habitual responses~\cite{wolpe2024mental}.
In \termName, one form of mental effort can be understood as the viewer's felt need to resist the impulse to stop before completion. This may involve resisting the temptation to abandon a long motion, or persisting despite physical fatigue and discomfort.
Similarly, \textit{cognitive effort} could serve as a visceral channel by having users complete cognitive tasks whose difficulty or duration increases with the \rv{information object's} size, similar to the \textit{United States of Abortion Mazes} article \cite{diehm2024united}, where readers are invited to solve one maze per state, each reflecting the strictness of its anti-abortion laws.

Generally speaking, visceral channels may be categorized as \textit{positive}, \textit{negative}, or \textit{ambiguous}, depending on how users generally experience them. While time and effort are generally ambiguous, discomfort is more clearly negative. Although negative visceral channels should be used with caution, their deliberate application may be worth examining in future work---for example, to amplify messages where the data already carries negative valence, such as pollution levels or pandemic mortality figures. This approach would ensure a form of semantic resonance \cite{lu2025designing}. In such cases, and with the user's informed consent, discomfort could be intensified to enhance its saliency as an encoding channel---for example, via induced motion sickness, uphill locomotion, or mild, safe forms of physical pain, anxiety, or deprivation (e.g., holding one's breath), experienced for longer durations as data values increase.

Conversely, positive data stories may be reinforced through positive visceral channels, where higher data values cause greater or longer-lasting pleasure. This can be achieved, for example, by linking the progressive \rv{examination of an information object} to enjoyable experiences such as a pleasant walk, appealing music, a pleasant scent \cite{patnaik2018information}, or a good meal \cite{datacuisine}.

\subsubsection{Variations around Progressive Value Reading}
It could be informative to investigate \termName beyond single users; for example, can \textit{joint \termName} (i.e., having multiple people participate in an operation, for example using a kids roundabout) promote engagement and understanding? Can \textit{vicarious \termName} (i.e., observing someone else perform \termName) enhance one's understanding of data? For example, the installation \textit{Of All The People In All The World} \cite{stan2004} represents statistics through piles of rice, with the artists continuously weighing small amounts and adding them by hand. \textit{Performative \termName} like this could permit forms that would be unethical for the general public but acceptable when devised and undertaken by an artist, such as Wafaa Bilal’s \cite{bilal2010counting} work, in which he had 15,000 war casualties tattooed on his back as individual dots during a 24-hour performance. Performance art may also allow viewers to witness \textit{prolonged \termName} operations that unfold over days, weeks, or even months, as a form of endurance art. Alternatively, prolonged \termName could occur through \textit{ritualistic \termName}, where an individual privately performs a long \termName task bit by bit each day over weeks or months, as a form of personal reflection---similar to spiritual practices involving slow constructions like sand mandalas \cite{roark2014sands}.

\subsection{Limitations}

Our goal was to propose an operational definition for \termName and a conceptual framework that distills key design elements, in order to support practical applications and lay the foundations for future research. Nonetheless, our design space has limitations. 
First, as we discussed in \cref{sec:dimension-iteration}, some design aspects (e.g., \rv{information object} types and motion characteristics) were not included because of a lack of reliable methods to extract the relevant parameters and label the cases. We also omitted general design strategies to promote engagement, such as the use of music and visual embellishments, because they are not specific to \termName, even though they can be effective in practice. 
Second, since the examples in our corpus are generally simple in design, we did not extensively explore how more advanced interaction techniques could support more complex designs. For example, \termName designs often involve off-screen content, and human–computer interaction research has investigated techniques to help users locate such content \cite{baudisch2003halo,gustafson2008wedge}.
Finally, our design space is descriptive rather than prescriptive, and the strategies it outlines still need empirical evaluation. 
% We hope it will facilitate future studies by providing a shared terminology and clarifying the key dimensions along which \termName designs can be compared.

% //////////////////////////////////////////
% MAYBE MENTION LATER / IN A DIFFERENT REVISION
% //////////////////////////////////////////

%\subsection{TEMPORARY --- Other things to discuss or not}
%\leni{The relationship between \termName and visualization scanning? --> the generalizability of \termName to different visualization designs.}

%\pierre{Moved from the body movements section:} Future studies can investigate how different body movements affect the cognition of viewers in \termName.

%\pierre{Mention somewhere that in examples such as the diving one, it can be ambiguous if it's the mark that's moving or the viewer.}

%\pierre{Collective experiences. For example shared handcar, tandem bikes, that merry-go-round propelled by bikes, dragon boat, Roundabout (play)}

%\pierre{Circular motions to save space.}

%% file: sections/7-conclusion.tex
\section{Conclusion}
Our conceptual framework—comprising the definition of \termName and the associated design space—is intended to support both researchers and practitioners. First, it provides a shared vocabulary: by clarifying what constitutes \termName and defining its key sub-concepts and dimensions, the framework offers terminology that can facilitate clearer communication in future studies and design work. Second, the design space can help structure thinking about designs making use of \termName. The three guiding questions—\textit{How is data represented?} \textit{How is motion achieved?} and \textit{What strategies support engagement and understanding?}—can assist designers in articulating options, considering constraints (such as data characteristics, audiences, and technical limitations), and reasoning about trade-offs. Third, the design space may help identify opportunities for innovation. Some dimensions include strategies illustrated only through hypothetical cases, suggesting areas that could be explored further, and the high-level strategies in \autoref{sec:general} can serve as starting points during ideation. Finally, the framework can support empirical work: because strategies within a dimension define comparable alternatives, the design space may assist researchers in constructing controlled comparisons and evaluating how different design choices influence user experience and understanding.

%% file: sections/8-appendix.tex
\section{appendix: \termName corpus}
\label{sec:appendix}
\noindent The following is the full list of \termName examples.
\begin{table}[!ht]
\centering
\footnotesize
\caption{The list of real-life examples of \termName}
\label{tab: caselist}
\begin{tabularx}{\textwidth}{@{} p{0.18cm} p{5cm} p{6cm} p{5cm} @{}}
\toprule
\textbf{ID} &\textbf{Case Title} &\textbf{Brief description} & \textbf{URL} \\
\midrule
C1 \label{ex:C1} &Wealth Shown To Scale &Long blocks, representing wealth of richest people in USA &\url{https://eattherichtextformat.github.io/1-pixel-wealth/} \\
\midrule 
C2 \label{ex:C2} &8 Billion People In Perspective &A 3D animation where the camera moves through the world population as a crowd &\url{https://www.youtube.com/watch?v=Sxejy6jd\_zc} \\
\midrule 
C3 \label{ex:C3} &Risk, Cycling And Denominator Neglect &Clusters of blocks representing bike journeys &\url{https://www.gicentre.net/blog/2013/11/24/risk-cycling-and-denominator-neglect} \\
\midrule 
C4 \label{ex:C4} &Yale'S \$32 Billion Endowment &A 200-meter-long strip painted on a street &\url{https://x.com/DavarianBaldwin/status/1388840756895526912} \\
\midrule 
C5 \label{ex:C5} &Des Dinosaures À L'Homme (From Dinosaurs To Humans) &An animated universe timeline in a monitor goes forward as the viewer turns the crank &\url{https://www.fleurdepapier.com/realisations/2015/musee-de-l-homme/des-dinosaures-a-l-homme/} \\
\midrule 
C6 \label{ex:C6} &The Historical Cost Of Light &Pixels representing the cost of light &\url{https://pudding.cool/2020/12/lighting-cost/} \\
\midrule
C7 \label{ex:C7} &Hiking Path Of The Solar System &A hiking trail where every meter corresponds to one million kilometers of actual distance. &\url{https://de.wikipedia.org/wiki/Planetenweg\_Uetliberg\#} \\
\midrule 
C8 \label{ex:C8} &Trail Of Time &Walkable path illustrating the time elapsed leading to the Great Canyon formation &\url{https://www.nps.gov/grca/planyourvisit/the-trail-of-time.htm} \\
\midrule 
C9 \label{ex:C9} &The Fallen Of World War II &Video moving along vertical bars representing the number of deaths &\url{https://www.youtube.com/watch?v=DwKPFT-RioU&t=340s} \\
\midrule 
C10 \label{ex:C10} &The Deep Sea &A visualization of animals living at different depth levels of the sea &\url{https://neal.fun/deep-sea/} \\
\midrule 
C11 \label{ex:C11} &The Actual Number Of Americans Jailed Or Imprisoned &Pictograms representing the actual number of Americans imprisoned &\url{https://mkorostoff.github.io/incarceration-in-real-numbers/} \\
\midrule 
C12 \label{ex:C12} &3D Bar Chart (VR 360 Video) & VR of 3D bar charts representing tea production per country. &\url{https://drive.google.com/file/d/10Uh-UaGk77gfJarSN0In-FlPuOEtlIoG/view?usp=sharing} \\
\midrule 
C13 \label{ex:C13} &The Incredible Depth Of Earth Layers &Animated video of a bar representing the depth of the Earth. &\url{https://www.youtube.com/watch?v=OzuKknkJt-I} \\
\midrule
C14 \label{ex:C14} &All People On One Page & Every individual in the current world population is represented by a pictogram. &\url{https://www.worldometers.info/watch/world-population/} \\
\midrule 
C15 \label{ex:C15} &Death In Syria &Dots representing killed people during the Syrian civil war. &\url{https://www.nytimes.com/interactive/2015/09/14/world/middleeast/syria-war-deaths.html} \\
\midrule 
C16 \label{ex:C16} &7 Billion People On 1 Page &Every individual in the current world population is represented by a pictogram &\url{https://www.7billionworld.com/} \\
\midrule 
C17 \label{ex:C17} &Du Papier Toilette (Oui Oui...) &Urolling a toilet paper roll to compare sizes of different digital media presented by the length of paper sheets &\url{https://limitesnumeriques.fr/sensibiliser/animation-numerique-responsable/demontage-conference} \\
\midrule 
C18 \label{ex:C18} &All Of The World’S Money And Markets In One Visualization &Visualization of squares representing the world's money &\url{https://www.visualcapitalist.com/all-of-the-worlds-money-and-markets-in-one-visualization-2020/} \\
\midrule 
C19 \label{ex:C19} &If The Moon Were One Pixel &Planets of the solar system is laid out left-to-right &\url{https://joshworth.com/dev/pixelspace/pixelspace\_solarsystem.html} \\
\midrule 
C20 \label{ex:C20} &O Cálculo De Uma Tragédia (The Calculation Of A Tragedy) &COVID-19 deaths in Brazil as dots stacked by day over time &\url{https://www.nexojornal.com.br/especial/2020/08/08/100-mil-mortes-no-brasil-o-calculo-de-uma-tragedia} \\
\bottomrule
\end{tabularx}
\end{table}

\begin{table}
\centering
\footnotesize
\caption{The list of real-life cases of \termName (continued)}
\begin{tabularx}{\textwidth}{@{} p{0.18cm} p{5cm} p{6cm} p{5cm} @{}}
\toprule
\textbf{ID} &\textbf{Case Title} &\textbf{Brief description} & \textbf{URL} \\
\midrule 
C21 \label{ex:C21} &What If I Told You: You Eat 3496 Litres Of Water &A pictogram of a wall of 15400 bottles &\url{https://thewaterweeat.com/} \\
\midrule 
C22 \label{ex:C22} &500,000 Lives Lost &A large scrollable bee-swarm plot, showing COVID-19 deaths over time &\url{https://www.reuters.com/graphics/HEALTH-CORONAVIRUS/USA-CASUALTIES-CHRONOLOGY/xklpyomnrpg/} \\
\midrule
C23 \label{ex:C23} &Sea Level Rise By The Numbers &Pictograms of Olympic-size swimming pools representing metlwater &\url{https://www.reuters.com/graphics/CLIMATECHANGE-GREENLAND-SEALEVELS/010080F60WR/} \\
\midrule
C24 \label{ex:C24} &Number Of Deaths IN WW2 Per Country &A 3D animation where the camera moves through 3D bars. &\url{https://www.youtube.com/watch?v=7cgRwDkP6vk}\\
\midrule
C25 \label{ex:C25} &US Debt Of \$20 Trillion Visualized In Stacks Of Physical Cash &A 3D animation where the camera moves through stacks of money&\url{https://www.youtube.com/watch?v=XqUwr-Nkq9g} \\
\midrule
C26 \label{ex:C26} &Ocean Depth Comparison &Video explaining how deep the ocean is by introducing several famous depth locations &\url{https://www.youtube.com/watch?v=Q5C7sqVe2Vg} \\
\midrule
C27 \label{ex:C27} &Turkey'S Toxic Dust &Zoom-out animation of an area representing damaged buildings after earthquake &\url{https://www.reuters.com/graphics/TURKEY-QUAKE/TOXINS/znvnbmyrzvl/} \\
\midrule
C28 \label{ex:C28} &Space Elevator &The visualization of the layers of the atmosphere &\url{https://neal.fun/space-elevator/} \\
\midrule
C29 \label{ex:C29} &The Data Physicalization Of Wealth Shown To Scale &Physical cubes present the wealth of normal and richest Americans &\url{https://www.youtube.com/watch?v=SktIkgYcFH0} \\
\midrule
C30 \label{ex:C30} &Climbingchart &An AR bar chart displayed vertically across several floors of a building &\url{https://hal.science/hal-05224602/document} \\
\midrule
C31 \label{ex:C31} &Walkingchart &An AR bar chart displayed horizontally over several meters of a parking lot &\url{https://hal.science/hal-05224602/document} \\
\midrule
C32 \label{ex:C32} &Augmented Reality Waste Data Representations &A augmented-reality pile of trash bags &\url{https://www.youtube.com/watch?v=c9EQ726EoUg} \\
\midrule
C33 \label{ex:C33} &Global Plastic Waste 1950-2019 & Animation of plastic bottles falling down &\url{https://www.youtube.com/watch?app=desktop&v=\_IcJlX-lOzM} \\
\midrule
C34 \label{ex:C34} &The Search For Answers &Visualization of plane wreckage at the bottom of oceans &\url{https://www.reuters.com/graphics/INDONESIA-CRASH/SEARCH/xegvbeywzpq/} \\
\midrule
C35 \label{ex:C35} &Sizing Up Australia’S Bushfires &Squares representing burnt area after Australian wildfire &\url{https://www.reuters.com/graphics/AUSTRALIA-BUSHFIRES-SCALE/0100B4VK2PN/} \\
\midrule
C36 \label{ex:C36} &Skyfall - Tracking The Asteroids That Menace Earth &Scrollable chart placing past and upcoming asteroid flybys along the Earth–Moon distance &\url{https://www.reuters.com/graphics/SPACE-ASTEROIDS/zdpxdnrzrpx/}\\
\bottomrule
\end{tabularx}
\end{table}

\begin{table}
\centering
\footnotesize
\caption{The list of hypothetical cases of \termName}
\begin{tabularx}{\textwidth}{@{} p{0.18cm} p{5cm} p{6cm} p{5cm} @{}}
\toprule
\textbf{ID} &\textbf{Case Title} &\textbf{Brief description} & \textbf{URL} \\\midrule
C37 \label{ex:C37} &Proxemics Interaction &2D-screen visualization showing bar charts based on the distance between the screen and the viewer &\url{https://progressive-value-reading.github.io/\#C37} \\
\midrule
C38 \label{ex:C38} &Take An Elevator &Viewers taking an elevator with a transparent window through which they can see a bar chart painted on the wall &\url{https://progressive-value-reading.github.io/\#C38} \\
\midrule
C39 \label{ex:C39} &Peephole Interaction &Viewers using a phone as a peephole view through which the viewer can inspect the visualization on a wall &\url{https://progressive-value-reading.github.io/\#C39} \\
\midrule
C40 \label{ex:C40} &Data Driven VR Roller Coaster &VR roller coaster trails consisting of concatenated stacks of 100-dollar bills, representing the wealth of the richest person in the world &\url{https://progressive-value-reading.github.io/\#C40} \\
\midrule
C41 \label{ex:C41} &VR Spaceship Traveling Through The Solar System &VR trip across the solar system to visualize distances between planets &\url{https://progressive-value-reading.github.io/\#C41} \\
\midrule
C42 \label{ex:C42} &Comparing Wealth In Stacks Of 100-Dollar Bills In AR &AR stacks of 100-dollar bills onto a street to compare the wealth of normal and the richest people. &\url{https://progressive-value-reading.github.io/\#C42} \\
\midrule
C43 \label{ex:C43} &Tissue Paper Box &Physical visualization of a tissues accumulation to visualize waste &\url{https://progressive-value-reading.github.io/\#C43} \\
\midrule
C44 \label{ex:C44} &VR Spaceship Circling The Earth Surrounded By Money &A virtual spaceship journey around the Earth that is surrounded by stacks of 100-dollar bills representing the wealth of the richest person &\url{https://progressive-value-reading.github.io/\#C44} \\
\midrule
C45 \label{ex:C45} &Biking Around The Planets in VR &Riding a stationary bike to circle down-scaled planets in virtual reality to compare their sizes &\url{https://progressive-value-reading.github.io/\#C45} \\
\midrule
C46 \label{ex:C46} &Queuing With All People In The World In VR &VR visualization of the world population all in a queue, moving forward each time an inhabitant quit the queue &\url{https://progressive-value-reading.github.io/\#C46} \\
\midrule
C47 \label{ex:C47} &Go Up Stairs And See The History &Vertical timeline representing humankind history accross several floors, requiring to climb stairs &\url{https://progressive-value-reading.github.io/\#C47} \\
\midrule
C48 \label{ex:C48} &Escape From A Room Where Money Fall From The Sky &Cumulative view of money filling a VR room &\url{https://progressive-value-reading.github.io/\#C48} \\
\midrule
C49 \label{ex:C49} &Data As Virtual Flow &Flow (e.g., sand or water) in AR &\url{https://progressive-value-reading.github.io/\#C49} \\
\midrule
C50 \label{ex:C50} &Convey Belt &Visualizations painted on a convey belt or physical information objects carried by a convey belt &\url{https://progressive-value-reading.github.io/\#C50} \\
\midrule
C51 \label{ex:C51} &Information Objects On Streetcars &Information Objects painted on streetcars (e.g., trams) move in and out of the field of viewers &\url{https://progressive-value-reading.github.io/\#C51} \\
\midrule
C52 \label{ex:C52} &Dragon Dance &Bar charts painted on a long giant puppet of a dragon in traditional Chinese dragon-dance &\url{https://progressive-value-reading.github.io/\#C52} \\
\midrule
C53 \label{ex:C53} &Visualization Shown On The Screen Of A Treadmill &Horizontal bar charts displayed on a 2D-screen moving forward by running on a treadmill &\url{https://progressive-value-reading.github.io/\#C53} \\
\midrule
C54 \label{ex:C54} &Moving Up An Elevator Through Fixed Pulleys In VR &Vertical bar charts displayed outside a VR elevator going up by pulling up fixed pulleys &\url{https://progressive-value-reading.github.io/\#C54} \\
\midrule
C55 \label{ex:C55} &Diving In VR &An VR experience where viewers mimic diving with a descent line to see how deep the sea is &\url{https://progressive-value-reading.github.io/\#C55} \\
\bottomrule
\end{tabularx}
\end{table}

%% file: ref.bib
@book{von2002language,
  title={The language of graphics: A framework for the analysis of syntax and meaning in maps, charts and diagrams},
  author={von Engelhardt, J{\"o}rg},
  year={2002},
  publisher={Yuri Engelhardt}
}

@inproceedings{yang2025making,
  title={Making Data Harder to Read: Visualizations that Intentionally Increase Time and Effort},
  author={Yang, Leni and Dragicevic, Pierre and Jansen, Yvonne},
  booktitle={IEEE Seventh Workshop on Visualization for Communication},
  pages={6--9},
  year={2025},
  organization={IEEE},
  doi="10.1109/VisComm69388.2025.00006"
}

@article{batziakoudi2024lost,
  title={Lost in Magnitudes: Exploring the Design Space for Visualizing Data with Large Value Ranges},
  author={Batziakoudi, Katerina and Cabric, Florent and Rey, St{\~A}{\v{S}}phanie and Fekete, Jean-Daniel},
  journal={arXiv preprint arXiv:2404.15150},
  year={2024},
  volume={1},
  number={1},
  pages={2426--2435},
doi = "https://doi.org/10.48550/arXiv.2404.15150"
}

@article{chevalier2013using,
  title={Using concrete scales: A practical framework for effective visual depiction of complex measures},
  author={Chevalier, Fanny and Vuillemot, Romain and Gali, Guia},
  journal={IEEE Transactions on Visualization and Computer Graphics},
  volume={19},
  number={12},
  pages={2426--2435},
  year={2013},
  publisher={IEEE},
doi = "https://doi.org/10.1109/tvcg.2013.210"
}

@inproceedings{lan2022negative,
  title={Negative emotions, positive outcomes? exploring the communication of negativity in serious data stories},
  author={Lan, Xingyu and Wu, Yanqiu and Shi, Yang and Chen, Qing and Cao, Nan},
  booktitle={Proceedings of the 2022 CHI conference on human factors in computing systems},
publisher = {Association for Computing Machinery},
address = {New York, NY, USA},
  pages={1--14},
  year={2022},
doi = "https://doi.org/10.1145/3491102.3517530"
}

@article{adar2020communicative,
  title={Communicative visualizations as a learning problem},
  author={Adar, Eytan and Lee, Elsie},
  journal={IEEE Transactions on Visualization and Computer Graphics},
  volume={27},
  number={2},
  pages={946--956},
  year={2020},
  publisher={IEEE},
doi = "https://doi.org/10.1109/TVCG.2020.3030375"
}

@article{lan2023affective,
  title={Affective visualization design: Leveraging the emotional impact of data},
  author={Lan, Xingyu and Wu, Yanqiu and Cao, Nan},
  journal={IEEE Transactions on Visualization and Computer Graphics},
  volume={30},
  number={1},
  pages={1--11},
  year={2023},
  publisher={IEEE}
}

@article{lee2022affective,
  title={Affective learning objectives for communicative visualizations},
  author={Lee-Robbins, Elsie and Adar, Eytan},
  journal={IEEE Transactions on Visualization and Computer Graphics},
  volume={29},
  number={1},
  pages={1--11},
  year={2022},
  publisher={IEEE},
doi = "https://doi.org/10.1109/TVCG.2022.3209500"
}

@inproceedings{chen2024beyond,
  title={Beyond numbers: Creating analogies to enhance data comprehension and communication with generative AI},
  author={Chen, Qing and Shuai, Wei and Zhang, Jiyao and Sun, Zhida and Cao, Nan},
  booktitle={Proceedings of the 2024 CHI Conference on Human Factors in Computing Systems},
  pages={1--14},
  year={2024},
doi = "https://doi.org/10.1145/3613904.3642480",
publisher = {Association for Computing Machinery},
address = {New York, NY, USA}
}

@article{lee2020data,
  title={Data visceralization: Enabling deeper understanding of data using virtual reality},
  author={Lee, Benjamin and Brown, Dave and Lee, Bongshin and Hurter, Christophe and Drucker, Steven and Dwyer, Tim},
  journal={IEEE Transactions on Visualization and Computer Graphics},
  volume={27},
  number={2},
  pages={1095--1105},
  year={2020},
  publisher={IEEE},
doi = "https://doi.org/10.1109/TVCG.2020.3030435"
}

@book{munzner2014visualization,
  title={Visualization analysis and design},
  author={Munzner, Tamara},
  year={2014},
  publisher={CRC press},
address = "Boca Raton, FL, USA"
}

@article{yao2022visualization,
  title={Visualization in motion: A research agenda and two evaluations},
  author={Yao, Lijie and Bezerianos, Anastasia and Vuillemot, Romain and Isenberg, Petra},
  journal={IEEE Transactions on Visualization and Computer Graphics},
  volume={28},
  number={10},
  pages={3546--3562},
  year={2022},
  publisher={IEEE},
doi = "https://doi.org/10.1109/TVCG.2022.3184993"
}

@article{solen2024designv1,
  title={A Design Space for Visualization with Large Scale-Item Ratios},
  author={Solen, Mara and Munzner, Tamara},
  journal={arXiv preprint arXiv:2404.01485v1},
  url={https://arxiv.org/pdf/2404.01485v1},
  year={2024},
pages="1-10",
volume = "1",
number = "1"
}

@article{solen2024designv2,
  title={A Design Space for Multiscale Visualization},
  author={Solen, Mara and Oddo, Matt and Munzner, Tamara},
  journal={IEEE Transactions on Visualization and Computer Graphics},
  url={https://arxiv.org/pdf/2404.01485v2},
  year={in press},
pages = "1-11",
volume = "1",
number = "1"
}

@article{segel2010narrative,
  title={Narrative visualization: Telling stories with data},
  author={Segel, Edward and Heer, Jeffrey},
  journal={IEEE Transactions on Visualization and Computer Graphics},
  volume={16},
  number={6},
  pages={1139--1148},
  year={2010},
  publisher={IEEE},
doi = "https://doi.org/10.1109/TVCG.2010.179"
}

@article{yang2021design,
  title={A design space for applying the freytag's pyramid structure to data stories},
  author={Yang, Leni and Xu, Xian and Lan, XingYu and Liu, Ziyan and Guo, Shunan and Shi, Yang and Qu, Huamin and Cao, Nan},
  journal={IEEE Transactions on Visualization and Computer Graphics},
  volume={28},
  number={1},
  pages={922--932},
  year={2021},
  publisher={IEEE},
doi = "https://doi.org/10.1109/TVCG.2021.3114774"
}

@inproceedings{amini2015understanding,
  title={Understanding data videos: Looking at narrative visualization through the cinematography lens},
  author={Amini, Fereshteh and Henry Riche, Nathalie and Lee, Bongshin and Hurter, Christophe and Irani, Pourang},
  booktitle={Proceedings of the 33rd Annual ACM conference on human factors in computing systems},
publisher = {Association for Computing Machinery},
address = {New York, NY, USA},
location = {Seoul, Korea},
  pages={1459--1468},
  year={2015},
doi="https://doi.org/10.1145/2702123.2702431"
}

@incollection{dragicevic2021data,
author="Dragicevic, Pierre
and Jansen, Yvonne
and Vande Moere, Andrew",
editor="Vanderdonckt, Jean
and Palanque, Philippe
and Winckler, Marco",
title="Data Physicalization",
bookTitle="Handbook of Human Computer Interaction",
year="2020",
publisher="Springer International Publishing",
address="Cham",
pages="1--51",
isbn="978-3-319-27648-9",
doi="10.1007/978-3-319-27648-9_94-1",
url="https://doi.org/10.1007/978-3-319-27648-9_94-1"
}

@article{hogan2017towards,
  title={Towards a design space for multisensory data representation},
  author={Hogan, Trevor and Hornecker, Eva},
  journal={Interacting with Computers},
  volume={29},
  number={2},
  pages={147--167},
  year={2017},
  publisher={Oxford University Press},
doi = "https://doi.org/10.1093/iwc/iww015"
}

@inproceedings{jansen2015opportunities,
  title={Opportunities and challenges for data physicalization},
  author={Jansen, Yvonne and Dragicevic, Pierre and Isenberg, Petra and Alexander, Jason and Karnik, Abhijit and Kildal, Johan and Subramanian, Sriram and Hornb{\ae}k, Kasper},
  booktitle={Proceedings of the 33rd annual acm conference on human factors in computing systems},
  pages={3227--3236},
  year={2015},
publisher = {Association for Computing Machinery},
address = {New York, NY, USA},
doi = "https://doi.org/10.1145/2702123.2702180"
}

@article{kaper1999data,
  title={Data sonification and sound visualization},
  author={Kaper, Hans G and Wiebel, Elizabeth and Tipei, Sever},
  journal={Computing in science \& engineering},
  volume={1},
  number={4},
  pages={48--58},
  year={1999},
  publisher={IEEE},
doi = "https://doi.org/10.1109/5992.774840"
}

@book{card1999readings,
  title={Readings in information visualization: using vision to think},
  author={Card, Stuart K and Mackinlay, Jock and Shneiderman, Ben},
  year={1999},
  publisher={Morgan Kaufmann},
address = {Burlington, MA, USA}
}

@article{hurtienne2020move,
  title={Move\&Find: The value of kinaesthetic experience in a casual data representation},
  author={Hurtienne, J{\"o}rn and Maas, Franzisca and Carolus, Astrid and Reinhardt, Daniel and Baur, Cordula and Wienrich, Carolin},
  journal={IEEE Computer Graphics and Applications},
  volume={40},
  number={6},
  pages={61--75},
  year={2020},
  publisher={IEEE},
doi = "https://doi.org/10.1109/mcg.2020.3025385"
}

@article{casamayou2022ride,
  title={Ride your data: Raise your arms, scream, and experience your data from a roller coaster cart},
  author={Casamayou, Vincent and Jansen, Yvonne and Dragicevic, Pierre and Prouzeau, Arnaud},
  journal={alt. VIS},
  volume={2},
  pages={3},
  year={2022},
url = "https://altvis.github.io/papers/2022/ride-your-data.pdf"
}

@article{hsu2018using,
  title={Using exaggerated feedback in a virtual reality environment to enhance behavior intention of water-conservation},
  author={Hsu, Wei-Che and Tseng, Ching-Mei and Kang, Shih-Chung},
  journal={Journal of Educational Technology \& Society},
  volume={21},
  number={4},
  pages={187--203},
  year={2018},
  publisher={JSTOR},
url = "https://www.jstor.org/stable/26511548"
}

@inproceedings{chauvergne2024weight,
  title={The Weight of Our Decisions: Encoding Carbon Impact with Physical Load},
  author={Chauvergne, Edwige and Ferron, Aymeric and Dragicevic, Pierre},
  booktitle={IHM 2024-35e Conf{\'e}rence Internationale Francophone sur l'Interaction Humain-Machine},
  year={2024},
  pages={1--8},
publisher = "HAL",
address = "Paris, France",
url = "https://hal.science/hal-04486642v1"
}

@article{slovic2007if,
  title={“If I look at the mass I will never act”: Psychic numbing and genocide},
  author={Slovic, Paul},
  journal={Judgment and Decision making},
  volume={2},
  number={2},
  pages={79--95},
  year={2007},
  publisher={Cambridge University Press},
doi = "10.1017/S1930297500000061"
}

@article{boyce2022large,
  title={Large numbers cause magnitude neglect: The case of government expenditures},
  author={Boyce-Jacino, Christina and Peters, Ellen and Galvani, Alison P and Chapman, Gretchen B},
  journal={Proceedings of the National Academy of Sciences},
  volume={119},
  number={28},
  pages={e2203037119},
  year={2022},
  publisher={National Academy of Sciences},
doi = "https://doi.org/10.1073/pnas.2203037119"
}

@inproceedings{veras2019saliency,
  title={Saliency deficit and motion outlier detection in animated scatterplots},
  author={Veras, Rafael and Collins, Christopher},
  booktitle={Proceedings of the 2019 CHI Conference on Human Factors in Computing Systems},
publisher = {Association for Computing Machinery},
address = {New York, NY, USA},
  pages={1--12},
  year={2019},
doi = "https://doi.org/10.1145/3290605.3300771"
}

@article{robertson2008effectiveness,
  title={Effectiveness of animation in trend visualization},
  author={Robertson, George and Fernandez, Roland and Fisher, Danyel and Lee, Bongshin and Stasko, John},
  journal={IEEE Transactions on Visualization and Computer Graphics},
  volume={14},
  number={6},
  pages={1325--1332},
  year={2008},
  publisher={IEEE},
doi = "https://doi.org/10.1109/TVCG.2008.125"
}

@article{jakobsen2013information,
  title={Information visualization and proxemics: Design opportunities and empirical findings},
  author={Jakobsen, Mikkel R and Haile, Yonas Sahlemariam and Knudsen, S{\o}ren and Hornb{\ae}k, Kasper},
  journal={IEEE Transactions on Visualization and Computer Graphics},
  volume={19},
  number={12},
  pages={2386--2395},
  year={2013},
  publisher={IEEE},
address = {New York, NY, USA},
doi = "https://doi.org/10.1109/TVCG.2013.166"
}

@article{isenberg2013hybrid,
  title={Hybrid-image visualization for large viewing environments.},
  author={Isenberg, Petra and Dragicevic, Pierre and Willett, Wesley and Bezerianos, Anastasia and Fekete, Jean-Daniel},
  journal={IEEE Trans. Vis. Comput. Graph.},
  volume={19},
  number={12},
  pages={2346--2355},
  year={2013},
doi = "http://dx.doi.org/10.1109/TVCG.2013.163"
}

@inproceedings{jansen2019effects,
  title={Effects of locomotion and visual overview on spatial memory when interacting with wall displays},
  author={Jansen, Yvonne and Schjerlund, Jonas and Hornb{\ae}k, Kasper},
  booktitle={Proceedings of the 2019 CHI Conference on Human Factors in Computing Systems},
publisher = {Association for Computing Machinery},
address = {New York, NY, USA},
  pages={1--12},
  year={2019},
doi = "https://doi.org/10.1145/3290605.3300521"
}

@inproceedings{jakobsen2015moving,
  title={Is moving improving? Some effects of locomotion in wall-display interaction},
  author={Jakobsen, Mikkel R and Hornb{\ae}k, Kasper},
  booktitle={Proceedings of the 33rd Annual ACM Conference on Human Factors in Computing Systems},
  pages={4169--4178},
  year={2015},
publisher = {Association for Computing Machinery},
address = {New York, NY, USA},
doi = "https://doi.org/10.1145/2702123.2702312"

}

@inproceedings{batziakoudi2024designing,
  title={Designing Visualizations for Enhancing Carbon Numeracy},
  author={Batziakoudi, Katerina and Cabric, Florent and Rey, St{\'e}phanie and Fekete, Jean-Daniel},
  booktitle={IEEE VIS 2024 Workshop on Visualization for Climate Action and Sustainability},
  year={2024},
publisher="HAL",
address = "Paris, France",
pages="1-4",
url = "https://inria.hal.science/hal-04744209v1"
}

@incollection{carson1997contingent,
  title={Contingent valuation surveys and tests of insensitivity to scope},
  author={Carson, Richard T},
  booktitle={Determining the value of non-marketed goods: economic, psychological, and policy relevant aspects of contingent valuation methods},
  pages={127--163},
  year={1997},
  publisher={Springer},
address = "New-York, NY, USA",
doi = "https://doi.org/10.1007/978-94-011-5364-5_6"
}

@book{desvousges2010measuring,
  title={Measuring nonuse damages using contingent valuation: An experimental evaluation of accuracy},
  author={Desvousges, William H and Johnson, F Reed and Dunford, Richard W and Boyle, Kevin J and Hudson, Sara P and Wilson, K Nicole},
  year={2010},
  publisher={RTI press},
address = "Research Triangle Park, NC, USA"
}

@article{morais2020showing,
  title={Showing data about people: A design space of anthropographics},
  author={Morais, Luiz and Jansen, Yvonne and Andrade, Nazareno and Dragicevic, Pierre},
  journal={IEEE Transactions on Visualization and Computer Graphics},
  volume={28},
  number={3},
  pages={1661--1679},
  year={2020},
  publisher={IEEE},
doi = "https://doi.org/10.1109/TVCG.2020.3023013"
}

@book{senay1990rules,
  title={Rules and principles of scientific data visualization},
  author={Senay, Hikmet and Ignatius, Eve},
  year={1990},
  publisher={Institute for Information Science and Technology, Department of Electrical~…},
address = "Washington, DC, USA"
}

@article{drucker2015unifying,
  title={A unifying framework for animated and interactive unit visualizations},
  author={Drucker, Steven and Fernandez, Roland},
  journal={Microsoft Research},
  volume={1},
  pages="1-9",
  year={2015}
}

@book{brinton1919graphic,
  title={Graphic methods for presenting facts},
  author={Brinton, Willard Cope},
  year={1919},
  publisher={Engineering magazine company},
address="New York, NY, USA"
}

@article{borgo2014order,
  title={Order of magnitude markers: An empirical study on large magnitude number detection},
  author={Borgo, Rita and Dearden, Joel and Jones, Mark W},
  journal={IEEE Transactions on Visualization and Computer Graphics},
  volume={20},
  number={12},
  pages={2261--2270},
  year={2014},
  publisher={IEEE},
doi = "https://doi.org/10.1109/tvcg.2014.2346428"
}

@article{isenberg2011study,
  title={A study on dual-scale data charts},
  author={Isenberg, Petra and Bezerianos, Anastasia and Dragicevic, Pierre and Fekete, Jean-Daniel},
  journal={IEEE Transactions on Visualization and Computer Graphics},
  volume={17},
  number={12},
  pages={2469--2478},
  year={2011},
  publisher={IEEE},
doi = "https://doi.org/10.1109/TVCG.2011.160"
}

@article{lan2021kineticharts,
  title={Kineticharts: Augmenting affective expressiveness of charts in data stories with animation design},
  author={Lan, Xingyu and Shi, Yang and Wu, Yanqiu and Jiao, Xiaohan and Cao, Nan},
  journal={IEEE Transactions on Visualization and Computer Graphics},
  volume={28},
  number={1},
  pages={933--943},
  year={2021},
  publisher={IEEE},
doi = "https://doi.org/10.1109/TVCG.2021.3114775"
}

@inproceedings{shi2021communicating,
  title={Communicating with motion: A design space for animated visual narratives in data videos},
  author={Shi, Yang and Lan, Xingyu and Li, Jingwen and Li, Zhaorui and Cao, Nan},
  booktitle={Proceedings of the 2021 CHI conference on human factors in computing systems},
  pages={1--13},
  year={2021},
publisher = {Association for Computing Machinery},
address = {New York, NY, USA},
doi = "https://doi.org/10.1145/3411764.3445337"
}

@inproceedings{amini2018hooked,
  title={Hooked on data videos: assessing the effect of animation and pictographs on viewer engagement},
  author={Amini, Fereshteh and Riche, Nathalie Henry and Lee, Bongshin and Leboe-McGowan, Jason and Irani, Pourang},
  booktitle={Proceedings of the 2018 International Conference on Advanced Visual Interfaces},
publisher = {Association for Computing Machinery},
address = {New York, NY, USA},
  pages={1--9},
  year={2018},
location = {Castiglione della Pescaia, Italy},
doi = "https://doi.org/10.1145/3206505.3206552"
}

@article{lu2020enhancing,
  title={Enhancing static charts with data-driven animations},
  author={Lu, Min and Fish, Noa and Wang, Shuaiqi and Lanir, Joel and Cohen-Or, Daniel and Huang, Hui},
  journal={IEEE Transactions on Visualization and Computer Graphics},
  volume={28},
  number={7},
  pages={2628--2640},
  year={2020},
  publisher={IEEE},
doi = "https://doi.org/10.1109/TVCG.2020.3037300"
}

@article{bartram2003moticons,
  title={Moticons:: detection, distraction and task},
  author={Bartram, Lyn and Ware, Colin and Calvert, Tom},
  journal={International Journal of Human-Computer Studies},
  volume={58},
  number={5},
  pages={515--545},
  year={2003},
  publisher={Elsevier},
doi = "https://doi.org/10.1016/S1071-5819(03)00021-1"
}

@article{bartram2002filtering,
  title={Filtering and brushing with motion},
  author={Bartram, Lyn and Ware, Colin},
  journal={Information Visualization},
  volume={1},
  number={1},
  pages={66--79},
  year={2002},
  publisher={Sage Publications Sage UK: London, England},
doi = "https://doi.org/10.1057/palgrave.ivs.9500005"
}

@article{ware2004motion,
  title={Motion to support rapid interactive queries on node--link diagrams},
  author={Ware, Colin and Bobrow, Robert},
  journal={ACM Transactions on Applied Perception (TAP)},
  volume={1},
  number={1},
  pages={3--18},
  year={2004},
  publisher={ACM New York, NY, USA},
doi = "https://doi.org/10.1145/1008722.1008724"
}

@inproceedings{bartram1997perceptual,
  title={Perceptual and interpretative properties of motion for information visualization},
  author={Bartram, Lyn},
  booktitle={Proceedings of the 1997 workshop on New paradigms in information visualization and manipulation},
  pages={3--7},
  year={1997},
doi="https://doi.org/10.1145/275519.275520",
publisher = {Association for Computing Machinery},
address = {New York, NY, USA},
location="Las Vegas, Nevada, USA"
}

@article{dimara2019interaction,
  title={What is interaction for data visualization?},
  author={Dimara, Evanthia and Perin, Charles},
  journal={IEEE Transactions on Visualization and Computer Graphics},
  volume={26},
  number={1},
  pages={119--129},
  year={2019},
  publisher={IEEE},
doi = "https://doi.org/10.1109/TVCG.2019.2934283"
}

@article{cockburn2009review,
  title={A review of overview+ detail, zooming, and focus+ context interfaces},
  author={Cockburn, Andy and Karlson, Amy and Bederson, Benjamin B},
  journal={ACM Computing Surveys (CSUR)},
  volume={41},
  number={1},
  pages={1--31},
  year={2009},
  publisher={ACM New York, NY, USA},
doi = "https://doi.org/10.1145/1456650.1456652"
}

@article{elmqvist2009melange,
  title={M{\'e}lange: Space folding for visual exploration},
  author={Elmqvist, Niklas and Riche, Yann and Henry-Riche, Nathalie and Fekete, Jean-Daniel},
  journal={IEEE Transactions on Visualization and Computer Graphics},
  volume={16},
  number={3},
  pages={468--483},
  year={2009},
  publisher={IEEE},
doi = "https://doi.org/10.1109/tvcg.2009.86"
}

@article{wolpe2024mental,
  title={What is mental effort: a clinical perspective},
  author={Wolpe, Noham and Holton, Richard and Fletcher, Paul C},
  journal={Biological Psychiatry},
  volume={95},
  number={11},
  pages={1030--1037},
  year={2024},
  publisher={Elsevier},
doi = "https://doi.org/10.1016/j.biopsych.2024.01.022"
}

@article{wiesing2023serial,
  title={Serial dependencies between locomotion and visual space},
  author={Wiesing, Michael and Zimmermann, Eckart},
  journal={Scientific Reports},
  volume={13},
  number={1},
  pages={3302},
  year={2023},
  publisher={Nature Publishing Group UK London},
doi = "https://doi.org/10.1038/s41598-023-30265-z"
}

@article{ruddle2011walking,
  title={Walking improves your cognitive map in environments that are large-scale and large in extent},
  author={Ruddle, Roy A and Volkova, Ekaterina and B{\"u}lthoff, Heinrich H},
  journal={ACM Transactions on Computer-Human Interaction (TOCHI)},
  volume={18},
  number={2},
  pages={1--20},
  year={2011},
  publisher={ACM New York, NY, USA},
doi = "https://doi.org/10.1145/1970378.1970384"
}

@article{clark1999embodied,
  title={An embodied cognitive science?},
  author={Clark, Andy},
  journal={Trends in cognitive sciences},
  volume={3},
  number={9},
  pages={345--351},
  year={1999},
  publisher={Elsevier},
doi = "https://doi.org/10.1016/S1364-6613(99)01361-3"
}

@article{duijzer2019embodied,
  title={Embodied learning environments for graphing motion: A systematic literature review},
  author={Duijzer, Carolien and Van den Heuvel-Panhuizen, Marja and Veldhuis, Michiel and Doorman, Michiel and Leseman, Paul},
  journal={Educational Psychology Review},
  volume={31},
  number={3},
  pages={597--629},
  year={2019},
  publisher={Springer},
doi = "https://doi.org/10.1007/s10648-019-09471-7"
}

@article{agostinho2015giving,
  title={Giving learning a helping hand: Finger tracing of temperature graphs on an iPad},
  author={Agostinho, Shirley and Tindall-Ford, Sharon and Ginns, Paul and Howard, Steven J and Leahy, Wayne and Paas, Fred},
  journal={Educational Psychology Review},
  volume={27},
  number={3},
  pages={427--443},
  year={2015},
  publisher={Springer},
doi = "https://doi.org/10.1007/s10648-015-9315-5"
}

@inproceedings{yang2023understanding,
  title={Understanding 3d data videos: From screens to virtual reality},
  author={Yang, Leni and Wu, Aoyu and Tong, Wai and Xu, Xian and Wei, Zheng and Qu, Huamin},
  booktitle={2023 IEEE 16th Pacific Visualization Symposium (PacificVis)},
  pages={197--206},
  year={2023},
  publisher={IEEE},
address = "New-York, NY, USA",
doi = "https://doi.org/10.1109/PacificVis56936.2023.00029"
}

@article{huron:hal-00846260,
  author={Huron, Samuel and Vuillemot, Romain and Fekete, Jean-Daniel},
  journal={IEEE Transactions on Visualization and Computer Graphics}, 
  title={Visual Sedimentation}, 
  year={2013},
  volume={19},
  number={12},
  pages={2446-2455},
  doi={10.1109/TVCG.2013.227}
}

@inproceedings{mackinlay1991perspective,
  title={The perspective wall: Detail and context smoothly integrated},
  author={Mackinlay, Jock D and Robertson, George G and Card, Stuart K},
  booktitle={Proceedings of the SIGCHI conference on Human factors in computing systems},
  pages={173--176},
  year={1991},
publisher = {Association for Computing Machinery},
address = {New York, NY, USA},
doi = "https://doi.org/10.1145/108844.108870"
}

@incollection{wilkinson2011grammar,
  title={The grammar of graphics},
  author={Wilkinson, Leland},
  booktitle={Handbook of computational statistics: Concepts and methods},
  pages={375--414},
  year={2011},
  publisher={Springer},
address = "New-York, NY, USA",
doi = "https://doi.org/10.1007/978-3-642-21551-3_13"
}

@article{satyanarayan2016vega,
  title={Vega-lite: A grammar of interactive graphics},
  author={Satyanarayan, Arvind and Moritz, Dominik and Wongsuphasawat, Kanit and Heer, Jeffrey},
  journal={IEEE Transactions on Visualization and Computer Graphics},
  volume={23},
  number={1},
  pages={341--350},
  year={2016},
  publisher={IEEE},
doi = "https://doi.org/10.1109/TVCG.2016.2599030"
}

@inproceedings{borgo2013glyph,
  title={Glyph-based Visualization: Foundations, Design Guidelines, Techniques and Applications.},
  author={Borgo, Rita and Kehrer, Johannes and Chung, David HS and Maguire, Eamonn and Laramee, Robert S and Hauser, Helwig and Ward, Matthew and Chen, Min},
  booktitle={Eurographics (state of the art reports)},
  pages={39--63},
  year={2013},
publisher = "Eurographics Association",
address="Eindhoven, The Netherlands ",
doi = "https://doi.org/10.2312/conf/EG2013/stars/039-063"
}

@inproceedings{haroz2015isotype,
  title={Isotype visualization: Working memory, performance, and engagement with pictographs},
  author={Haroz, Steve and Kosara, Robert and Franconeri, Steven L},
  booktitle={Proceedings of the 33rd annual ACM conference on human factors in computing systems},
  pages={1191--1200},
  year={2015},
publisher = {Association for Computing Machinery},
address = {New York, NY, USA},
doi = "https://doi.org/10.1145/2702123.2702275"
}

@article{patnaik2018information,
  title={Information olfactation: Harnessing scent to convey data},
  author={Patnaik, Biswaksen and Batch, Andrea and Elmqvist, Niklas},
  journal={IEEE Transactions on Visualization and Computer Graphics},
  volume={25},
  number={1},
  pages={726--736},
  year={2018},
  publisher={IEEE},
doi = "https://doi.org/10.1109/TVCG.2018.2865237"
}

@article{kahneman1999economic,
  title={Economic preferences or attitude expressions?: an analysis of dollar responses to public issues},
  author={Kahneman, Daniel and Ritov, Ilana and Schkade, David},
  journal={Journal of Risk and Uncertainty},
  volume={19},
  number={1},
  pages={203--235},
  year={1999},
  publisher={Springer},
doi = "https://doi.org/10.1023/A:1007835629236"
}

@inproceedings{asif2010exploring,
  title={Exploring distance encodings with a tactile display to convey turn by turn information in automobiles},
  author={Asif, Amna and Heuten, Wilko and Boll, Susanne},
  booktitle={Proceedings of the 6th Nordic Conference on Human-computer Interaction: Extending boundaries},
  pages={32--41},
  year={2010},
doi="https://doi.org/10.1145/1868914.1868923",
publisher = {Association for Computing Machinery},
address = {New York, NY, USA},
location="Reykjavik, Iceland"
}

@inproceedings{ferguson2018evaluating,
  title={Evaluating mapping designs for conveying data through tactons},
  author={Ferguson, Jamie and Williamson, John and Brewster, Stephen},
  booktitle={Proceedings of the 10th Nordic Conference on Human-Computer Interaction},
  pages={215--223},
  year={2018},
publisher = {Association for Computing Machinery},
address = {New York, NY, USA},
doi = "https://doi.org/10.1145/3240167.3240175"
}

@phdthesis{elmquist2025sensibly,
  title     = {Sensibly Sound: Human-Centered Integration of Sonification and Visualization},
  author    = {Elmquist, Elias},
  school    = {Link{\"o}ping University Electronic Press},
  year      = {2025},
}

@inproceedings{smith1991data,
  title={Data sonification: issues and challenges},
  author={Smith, Stuart},
  booktitle={Final Program and Paper Summaries 1991 IEEE ASSP Workshop on Applications of Signal Processing to Audio and Acoustics},
  pages={0\_63--0\_65},
  year={1991},
  publisher={IEEE},
address = "New-York, NY, USA",
doi = "https://doi.org/10.1109/ASPAA.1991.634106"
}

@inproceedings{chevalier2016animations,
  title={Animations 25 years later: New roles and opportunities},
  author={Chevalier, Fanny and Riche, Nathalie Henry and Plaisant, Catherine and Chalbi, Amira and Hurter, Christophe},
  booktitle={Proceedings of the international working conference on advanced visual interfaces},
  pages={280--287},
  year={2016},
publisher = {Association for Computing Machinery},
address = {New York, NY, USA},
doi = "http://dx.doi.org/10.1145/2909132.2909255"
}

@article{heer2007animated,
  title={Animated transitions in statistical data graphics},
  author={Heer, Jeffrey and Robertson, George},
  journal={IEEE transactions on visualization and computer graphics},
  volume={13},
  number={6},
  pages={1240--1247},
  year={2007},
  publisher={IEEE},
doi = "https://doi.org/10.1109/tvcg.2007.70539"
}

@article{rosling2011health,
  title={Health advocacy with Gapminder animated statistics},
  author={Rosling, Hans and Zhang, Zhongxing},
  journal={Journal of epidemiology and global health},
  volume={1},
  number={1},
  pages={11--14},
  year={2011},
  publisher={Elsevier},
doi = "https://doi.org/10.1016/j.jegh.2011.07.001"
}

@article{paneels2009review,
  title={Review of designs for haptic data visualization},
  author={Paneels, Sabrina and Roberts, Jonathan C},
  journal={IEEE Transactions on Haptics},
  volume={3},
  number={2},
  pages={119--137},
  year={2009},
  publisher={IEEE},
doi = "https://doi.org/10.1109/TOH.2009.44"
}

@inproceedings{seyser2018scrollytelling,
  title={Scrollytelling--an analysis of visual storytelling in online journalism},
  author={Seyser, Doris and Zeiller, Michael},
  booktitle={2018 22nd international conference information visualisation (IV)},
  pages={401--406},
  year={2018},
  publisher={IEEE},
address = "New-York, NY, USA",
doi = "https://doi.org/10.1109/iV.2018.00075"
}

@inproceedings{tjarnhage2023impact,
  title={The impact of scrollytelling on the reading experience of long-form journalism},
  author={Tj{\"a}rnhage, Anja and S{\"o}derstr{\"o}m, Ulrik and Norberg, Ole and Andersson, Mattias and Mejtoft, Thomas},
  booktitle={Proceedings of the European Conference on Cognitive Ergonomics 2023},
  pages={1--9},
  year={2023},
publisher = {Association for Computing Machinery},
address = {New York, NY, USA},
doi = "https://doi.org/10.1145/3605655.3605683"
}

@inproceedings{ball2007move,
author = {Ball, Robert and North, Chris and Bowman, Doug A.},
title = {Move to improve: promoting physical navigation to increase user performance with large displays},
year = {2007},
isbn = {9781595935939},
publisher = {Association for Computing Machinery},
address = {New York, NY, USA},
url = {https://doi.org/10.1145/1240624.1240656},
doi = {10.1145/1240624.1240656},
booktitle = {Proceedings of the SIGCHI Conference on Human Factors in Computing Systems},
pages = {191–200},
numpages = {10},
location = {San Jose, California, USA},
series = {CHI '07}
}

@incollection{peng2024telling,
  title={Telling a Story with Graphs: How to Apply Narrative Visualization Strategies and Visualization Techniques in Journalistic Storytelling},
  author={Peng, Zhao},
  booktitle={The Palgrave Handbook of Global Digital Journalism},
  pages={283--306},
  year={2024},
  publisher={Springer},
address = "New-York, NY, USA",
doi = "https://doi.org/10.1007/978-3-031-59379-6_18"
}

@inproceedings{jain2025strollytelling,
author = {Jain, Radhika Pankaj and Drogemuller, Adam and Satriadi, Kadek Ananta and Smith, Ross and Cunningham, Andrew},
title = {Strollytelling: Coupling Animation with Physical Locomotion to Explore Immersive Data Stories},
year = {2025},
isbn = {9798400713941},
publisher = {Association for Computing Machinery},
address = {New York, NY, USA},
url = {https://doi.org/10.1145/3706598.3713132},
doi = {10.1145/3706598.3713132},
booktitle = {Proceedings of the 2025 CHI Conference on Human Factors in Computing Systems},
articleno = {1006},
numpages = {17},
location = {
},
series = {CHI '25}
}

@article{fekete2016progressive,
  title={Progressive analytics: A computation paradigm for exploratory data analysis},
  author={Fekete, Jean-Daniel and Primet, Romain},
  journal={arXiv preprint arXiv:1607.05162},
  year={2016},
pages="1-10",
number = "1",
volume="1",
doi = "https://doi.org/10.48550/arXiv.1607.05162"
}

@article{hullman2011benefitting,
  title={Benefitting infovis with visual difficulties},
  author={Hullman, Jessica and Adar, Eytan and Shah, Priti},
  journal={IEEE Transactions on Visualization and Computer Graphics},
  volume={17},
  number={12},
  pages={2213--2222},
  year={2011},
  publisher={IEEE},
doi = "https://doi.org/10.1109/tvcg.2011.175"
}

@article{palmisano2015future,
  title={Future challenges for vection research: definitions, functional significance, measures, and neural bases},
  author={Palmisano, Stephen and Allison, Robert S and Schira, Mark M and Barry, Robert J},
  journal={Frontiers in psychology},
  volume={6},
  pages={193},
  year={2015},
  publisher={Frontiers Media SA},
doi = "https://doi.org/10.3389/fpsyg.2015.00193"
}

@article{block2016physical,
  title={Physical load affects duration judgments: A meta-analytic review},
  author={Block, Richard A and Hancock, PA and Zakay, Dan},
  journal={Acta psychologica},
  volume={165},
  pages={43--47},
  year={2016},
  publisher={Elsevier},
doi = "https://doi.org/10.1016/j.actpsy.2016.01.002"
}

@article{yanez2025state,
  title={The State of the Art in User-Adaptive Visualizations},
  author={Yanez, Fernando and Conati, Cristina and Ottley, Alvitta and Nobre, Carolina},
  journal={Computer Graphics Forum},
  volume={44},
  number={1},
  pages={e15271},
  year={2025},
  organization={Wiley Online Library},
  doi = "https://doi.org/10.1111/cgf.15271"
}

@article{lopes2022perceived,
  title={Perceived exertion: Revisiting the history and updating the neurophysiology and the practical applications},
  author={Lopes, Thiago Ribeiro and Pereira, Hugo Maxwell and Silva, Bruno Moreira},
  journal={International Journal of Environmental Research and Public Health},
  volume={19},
  number={21},
  pages={14439},
  year={2022},
  publisher={MDPI},
doi = "https://doi.org/10.3390/ijerph192114439"
}

@article{zinoubi2018relationships,
  title={Relationships between rating of perceived exertion, heart rate and blood lactate during continuous and alternated-intensity cycling exercises},
  author={Zinoubi, Badrane and Zbidi, Sana and Vandewalle, Henry and Chamari, Karim and Driss, Tarak},
  journal={Biology of sport},
  volume={35},
  number={1},
  pages={29--37},
  year={2018},
  publisher={Termedia},
doi = "https://doi.org/10.5114/biolsport.2018.70749"
}

@article{faulkner2008rating,
  title={The rating of perceived exertion during competitive running scales with time},
  author={Faulkner, James and Parfitt, Gaynor and Eston, Roger},
  journal={Psychophysiology},
  volume={45},
  number={6},
  pages={977--985},
  year={2008},
  publisher={Wiley Online Library},
doi = "https://doi.org/10.1111/j.1469-8986.2008.00712.x"
}

@article{borg1987perceived,
  title={Perceived exertion related to heart rate and blood lactate during arm and leg exercise},
  author={Borg, Gunnar and Hassm{\'e}n, Peter and Lagerstr{\"o}m, Monica},
  journal={European journal of applied physiology and occupational physiology},
  volume={56},
  number={6},
  pages={679--685},
  year={1987},
  publisher={Springer},
doi = "https://doi.org/10.1007/BF00424810"
}

@article{green2004heart,
  title={Heart rate and ratings of perceived exertion during treadmill and elliptical exercise training},
  author={Green, James M and Crews, Thad R and Pritchett, Robert C and Mathfield, Chaye and Hall, Laura},
  journal={Perceptual and motor skills},
  volume={98},
  number={1},
  pages={340--348},
  year={2004},
  publisher={SAGE Publications Sage CA: Los Angeles, CA},
doi = "https://doi.org/10.2466/pms.98.1.340-348"
}

@article{cleveland_experiment_1986,
	title = {An experiment in graphical perception},
	volume = {25},
	issn = {0020-7373},
	url = {http://www.sciencedirect.com/science/article/pii/S0020737386800190},
	doi = {10.1016/S0020-7373(86)80019-0},
	number = {5},
	journal = {Int. J. Man. Mach. Stud.},
	author = {Cleveland, William S and McGill, Robert},
	month = nov,
	year = {1986},
	pages = {491--500},
}

@article{hollands_judgments_1992,
	title = {Judgments of change and proportion in graphical perception},
	volume = {34},
	issn = {0018-7208},
	url = {http://www.ncbi.nlm.nih.gov/pubmed/1634243},
	number = {3},
	journal = {Hum. Factors},
	author = {Hollands, J G and Spence, I},
	month = jun,
	year = {1992},
	pages = {313--334},
}

@article{talbot_four_2014,
	title = {Four {Experiments} on the {Perception} of {Bar} {Charts}},
	volume = {20},
	issn = {1077-2626},
	url = {http://dx.doi.org/10.1109/TVCG.2014.2346320},
	doi = {10.1109/TVCG.2014.2346320},
	number = {12},
	journal = {IEEE Trans. Vis. Comput. Graph.},
	author = {Talbot, J and Setlur, V and Anand, A},
	month = dec,
	year = {2014},
	pages = {2152--2160},
}

@article{legge_efficiency_1989,
	title = {Efficiency of graphical perception},
	volume = {46},
	issn = {0031-5117},
	doi = {10.3758/bf03204990},
	language = {eng},
	number = {4},
	journal = {Perception \& Psychophysics},
	author = {Legge, G. E. and Gu, Y. C. and Luebker, A.},
	month = oct,
	year = {1989},
	pmid = {2798030},
	pages = {365--374},
}

@article{banton_perception_2005,
	title = {The {Perception} of {Walking} {Speed} in a {Virtual} {Environment}},
	volume = {14},
	issn = {1054-7460},
	url = {https://direct.mit.edu/pvar/article/14/4/394-406/18584},
	doi = {10.1162/105474605774785262},
	language = {en},
	number = {4},
	urldate = {2025-09-05},
	journal = {Presence: Teleoperators and Virtual Environments},
	author = {Banton, Tom and Stefanucci, Jeanine and Durgin, Frank and Fass, Adam and Proffitt, Dennis},
	month = aug,
	year = {2005},
	pages = {394--406},
}

@article{izard_calibrating_2008,
	title = {Calibrating the mental number line},
	volume = {106},
	issn = {0010-0277},
	url = {https://www.sciencedirect.com/science/article/pii/S0010027707001564},
	doi = {10.1016/j.cognition.2007.06.004},
	number = {3},
	urldate = {2025-09-07},
	journal = {Cognition},
	author = {Izard, Véronique and Dehaene, Stanislas},
	month = mar,
	year = {2008},
	pages = {1221--1247},
}

@book{dehaene_number_1997,
	address = {New York},
	title = {The number sense: how the mind creates mathematics},
	isbn = {978-0-19-511004-3},
	shorttitle = {The number sense},
	publisher = {Oxford University Press},
	author = {Dehaene, Stanislas},
	year = {1997},
}

@article{leibovich_sense_2017,
	title = {From “sense of number” to “sense of magnitude”: {The} role of continuous magnitudes in numerical cognition},
	volume = {40},
	copyright = {https://www.cambridge.org/core/terms},
	issn = {0140-525X, 1469-1825},
	shorttitle = {From “sense of number” to “sense of magnitude”},
	url = {https://www.cambridge.org/core/product/identifier/S0140525X16000960/type/journal_article},
	doi = {10.1017/S0140525X16000960},
	language = {en},
	urldate = {2025-09-07},
	journal = {Behavioral and Brain Sciences},
	author = {Leibovich, Tali and Katzin, Naama and Harel, Maayan and Henik, Avishai},
	year = {2017},
	pages = {e164},
}

@inproceedings{Correll2014Brain,
  year={2014},
  title={Bad for Data, Good for the Brain : Knowledge-First Axioms For Visualization Design},
  booktitle={DECISIVe : Workshop on Dealing with Cognitive Biases in Visualisations},
  editor={Ellis, Geoffrey},
  author={Correll, Michael and Gleicher, Michael},
publisher = {IEEE},
address = {New York, NY, USA},
pages="1-6",
url = "http://nbn-resolving.de/urn:nbn:de:bsz:352-0-329455"
}

@inproceedings{bertini2020shouldn,
  title={Why shouldn’t all charts be scatter plots? beyond precision-driven visualizations},
  author={Bertini, Enrico and Correll, Michael and Franconeri, Steven},
  booktitle={Proceedings of the 2020 IEEE Visualization Conference},
  pages={206--210},
  year={2020},
  publisher={IEEE},
address = {New York, NY, USA},
doi="https://doi.org/10.1109/VIS47514.2020.00048"
}

@misc{debur2005Haring,
author = {Emmanuelle Debur},  
title = {À Bordeaux, une œuvre de Keith Haring de 10 mètres dans l’ascenseur du CAPC},
year = 2025,
  howpublished = {Sud Ouest. \url{https://www.sudouest.fr/gironde/bordeaux/video-a-bordeaux-une-oeuvre-de-keith-haring-de-10-metres-dans-l-ascenseur-du-capc-25369111.php}},
  note = {Accessed: 2025-09-08}
}

@inproceedings{Ferron.etal2025,
  title = {{{AROM}}: {{Rambling Along Data}} in {{Augmented Reality}} to {{Explore Large Order}} of {{Magnitude Values}}},
  shorttitle = {{{AROM}}},
year = "2025",
  booktitle = {Workshop on {{Visualization}} for {{Communication}} at the 2025 {{IEEE Visualization Conference}}},
  author = {Ferron, Aymeric and Jansen, Yvonne and Hachet, Martin},
  date = {2025-11},
  location = {Vienne, Austria},
  url = {https://hal.science/hal-05224602},
  urldate = {2025-09-08},
pages = "1-6",
publisher = {IEEE},
address = {New York, NY, USA},
}

@article{kosmas2019implementing,
  title={Implementing embodied learning in the classroom: Effects on children’s memory and language skills},
  author={Kosmas, Panagiotis and Ioannou, Andri and Zaphiris, Panayiotis},
  journal={Educational Media International},
  volume={56},
  number={1},
  pages={59--74},
  year={2019},
  publisher={Taylor \& Francis},
doi = "https://doi.org/10.1080/09523987.2018.1547948"
}

@article{pousman2007casual,
  title={Casual information visualization: Depictions of data in everyday life},
  author={Pousman, Zachary and Stasko, John and Mateas, Michael},
  journal={IEEE transactions on visualization and computer graphics},
  volume={13},
  number={6},
  pages={1145--1152},
  year={2007},
  publisher={IEEE},
doi = "https://doi.org/10.1109/TVCG.2007.70541"
}

@inproceedings{baudisch2003halo,
  title={Halo: a technique for visualizing off-screen objects},
  author={Baudisch, Patrick and Rosenholtz, Ruth},
  booktitle={Proceedings of the SIGCHI conference on Human factors in computing systems},
  pages={481--488},
  year={2003},
publisher = {Association for Computing Machinery},
address = {New York, NY, USA},
doi="https://doi.org/10.1145/642611.642695"
}

@book{noe2004action,
  title={Action in perception},
  author={No{\"e}, Alva},
  year={2004},
  publisher={MIT press},
address = "Cambridge, MA, USA"
}

@inproceedings{petiot2025effect,
  title={The Effect of Augmented Reality on Involuntary Autobiographical Memory},
  author={Petiot, L{\'e}ana and Sauz{\'e}on, H{\'e}l{\`e}ne and Dragicevic, Pierre},
  booktitle={Proceedings of the 2025 CHI Conference on Human Factors in Computing Systems},
  pages={1--20},
  year={2025},
publisher = {Association for Computing Machinery},
address = {New York, NY, USA},
doi = "https://doi.org/10.1145/3706598.3713922"
}

@article{norcio1989adaptive,
  title={Adaptive human-computer interfaces: A literature survey and perspective},
  author={Norcio, Anthony F and Stanley, Jaki},
  journal={IEEE Transactions on Systems, Man, and cybernetics},
  volume={19},
  number={2},
  pages={399--408},
  year={1989},
  publisher={IEEE},
doi = "https://doi.org/10.1109/21.31042"
}

@article{lu2025designing,
  title={Designing Semantically-Resonant Abstract Patterns for Data Visualization},
  author={Lu, Zihan and He, Tingying and Hong, Jiayi and Yao, Lijie and Isenberg, Tobias},
  journal={arXiv preprint arXiv:2505.14816},
  year={2025},
volume = "1",
number = "1",
pages = "1-10",
doi = "https://doi.org/10.48550/arXiv.2505.14816"
}

@misc{diehm2024united,
author = {Jan Diehm, Michelle Pera-McGhee},  
title = {United States of Abortion Mazes},
year = 2024,
  howpublished = {The Pudding. \url{https://pudding.cool/2024/10/abortion-mazes/}},
  note = {Accessed: 2025-09-11}
}

@misc{datacuisine,
author = {Suzanne Jaschko and Moritz Stefaner},  
title = {Data Cuisine},
year = {2012--2025},
  howpublished = {\url{https://data-cuisine.net/}},
  note = {Accessed: 2025-09-11}
}

@misc{stan2004,
author = {Stan's Cafe},  
title = {Of All The People In All The World},
year = {2004},
  howpublished = {\url{https://stans.cafe/project/project-of-all-the-people/}},
  note = {Accessed: 2025-09-11}
}

@article{roark2014sands,
  title={The Sands of Performance},
  author={Roark, Carolyn},
  journal={Prabuddha Bharata or Awakened India},
  volume={119},
    url={https://www.esamskriti.com/essays/pdf/14-aug-The-Sands-of-Performance.pdf},
    year={2014},
  pages={478--486}
}

@misc{bilal2010counting,
  title={and Counting...},
  author={Wafaa Bilal},
year = {2010},
  howpublished = {\url{https://wafaabilal.com/and-counting/}},
  note = {Accessed: 2025-09-11}
}

@inproceedings{gustafson2008wedge,
  title={Wedge: clutter-free visualization of off-screen locations},
  author={Gustafson, Sean and Baudisch, Patrick and Gutwin, Carl and Irani, Pourang},
  booktitle={Proceedings of the SIGCHI conference on human factors in computing systems},
  pages={787--796},
  year={2008},
publisher = {Association for Computing Machinery},
address = {New York, NY, USA},
doi = "https://doi.org/10.1145/1357054.1357179"
}

@article{batziakoudi2026beyond,
  TITLE = {{Beyond Log Scales: Toward Cognitively Informed Bar Charts for Orders of Magnitude Values}},
  AUTHOR = {Batziakoudi, Katerina and Rey, St{\'e}phanie and Fekete, Jean-Daniel},
  URL = {https://hal.science/hal-05171203},
  JOURNAL = {{IEEE Transactions on Visualization and Computer Graphics}},
  PUBLISHER = {{Institute of Electrical and Electronics Engineers}},
  YEAR = {2026},
  HAL_ID = {hal-05171203},
  HAL_VERSION = {v1},
}

@article{ulmer2023survey,
  title={A survey on progressive visualization},
  author={Ulmer, Alex and Angelini, Marco and Fekete, Jean-Daniel and Kohlhammer, J{\"o}rn and May, Thorsten},
  journal={IEEE Transactions on Visualization and Computer Graphics},
  volume={30},
  number={9},
  pages={6447--6467},
  year={2023},
  publisher={IEEE}
}
